\documentclass[10pt,twocolumn,letterpaper]{article}

\usepackage{iccv}
\usepackage{times}
\usepackage{epsfig}
\usepackage{graphicx}
\usepackage{amsmath}
\usepackage{amssymb}
\usepackage{overpic}
\usepackage{xcolor}
\usepackage{bm}
\usepackage{subcaption}
\usepackage[symbol]{footmisc}

\usepackage[breaklinks=true,bookmarks=false]{hyperref}

\iccvfinalcopy 

\newcommand{\jmb}[1]{{\bm{#1}}}
\newcommand{\jcb}[1]{\pmb{#1}}

\begin{document}

\title{DSIC: Deep Stereo Image Compression}

\author{Jerry Liu \footnotemark[3] , Shenlong Wang, and Raquel Urtasun\\
Uber ATG\\
{\tt\small jerryl@uber.com, slwang@uber.com, urtasun@uber.com}
}

\maketitle

\begin{abstract}

In this paper we tackle the problem of stereo image compression, and  leverage the fact that  the two images have overlapping fields of view to further compress the representations. 
Our approach leverages  state-of-the-art single-image compression autoencoders and enhances the compression  with  novel  \textit{parametric skip functions} to feed fully differentiable, disparity-warped features at all levels to the encoder/decoder of the second image. Moreover, we model the probabilistic dependence between the image codes using a \textit{conditional entropy model}.
Our experiments show an impressive 
$30\operatorname{-}50\%$
reduction in the second image bitrate at low bitrates compared to deep single-image compression, and a 
$10\operatorname{-}20\%$
reduction at higher bitrates.

\end{abstract}

\footnotetext[3]{Work done as part of the Uber AI Residency program.}
\section{Introduction}

Many applications such as autonomous vehicles and 3D movies involve the use of stereo camera pairs. These arrays of cameras oftentimes capture and store massive quantities of data per day, which require good image compression algorithms to ensure an efficient use of space. A naive approach to image compression would be to compress the image streams from each camera separately. However, this ignores the shared information given by the overlapping fields of view between the cameras. 
Hence, there is a need for compression methods that can efficiently compress a stereo image pair further by utilizing the shared information. 

Stereo image compression can be seen as in-between the work of image and video compression. While we get to utilize shared information between two images, we are not able to exploit the spatial-temporal redundancies within a tightly coupled image sequence. 
There has been an abundance of work on traditional multi-view and stereo compression \cite{flierl_disparitymvc, frajka_resstereo} as well as  deep-learning based image and video compression \cite{balle_varhyperprior, minnen_jointpriors, rippel_learnedvidcomp, wu_vidinterpolation}. However, the space of deep multi-view compression is relatively unexplored. 

In this work, we present a novel end-to-end deep architecture for stereo image compression. Our contributions revolve around trying to extract as much information out of the first image in order to reduce the bitrate of the second. Towards this goal we leverage state-of-the-art single-image compression autoencoders, and enhance them with  novel  \textit{parametric skip functions} to feed fully differentiable, disparity-warped features at all levels to the encoder/decoder of the second image. This allow us to store fewer bits for the second image code since multi-level information is being passed from the encoder/decoder of the first image. 
Moreover, we model the probabilistic dependence between the image codes using a \textit{conditional entropy model}. Since the codes of a stereo pair are highly correlated with each other, a model that can capture this dependence will help reduce the joint entropy, and hence the joint bitrate, of the two latent codes.  

We demonstrate a 
$30\operatorname{-}50\%$
 reduction in the second image bitrate at low bitrates compared to deep single-image compression, and a 
$10\operatorname{-}20\%$
 reduction at higher bitrates, when evaluated over an internal self-driving dataset (NorthAmerica), as well as stereo images from Cityscapes.
Our experiments additionally demonstrate that we outperform all image codecs and motion-compensation+residual coding baselines on MS-SSIM, a perceptual metric capturing the structural quality of an image. 
\section{Background and Related Work}
\begin{figure*}[t]
	\centering
	\includegraphics[width=.9\linewidth]{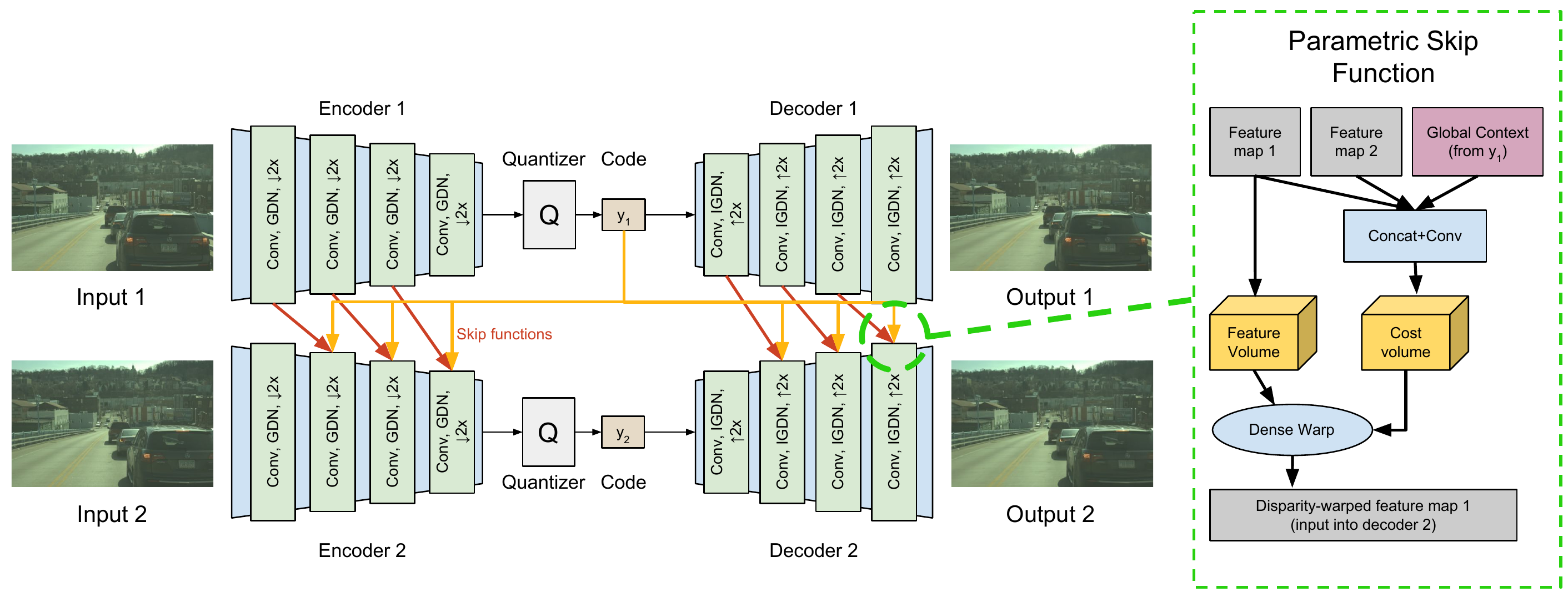}
		\vspace{-3mm}
	\caption{Left: the end-to-end stereo compression architecture; right: the proposed deep parameteric skip function. 
	}
	\label{fig:stereo_complex_model}
	\vspace{-3mm}
\end{figure*}

We start this section with a brief overview of deep image compression algorithms, including the general formulation and previous works. We then discuss related work on deep video compression, multi-view compression and stereo estimation, respectively.
\subsection{A Review on Deep Image Compression} 
There has been a plethora of work on learned, single-image, lossy image compression \cite{toderici_varimgcomp, toderici_fullimgcomp, balle_imgcomp_end2end, balle_varhyperprior, theis_imgcomp_ae, mentzer_condprobimg, rippel_rtimgcomp, minnen_jointpriors, lee_contextadapt}. These works generally use nonlinear transforms through convolutional neural network (CNN) layers to encode an image into a latent space, which is then quantized  into discrete symbols. An entropy coding function using a learned entropy model is then applied to losslessly produce the final bitstream. 

More formally, given the input $\jmb{x}$, deep image compression learns an encoding+quantization function $E(\cdot)$  mapping the input to a discrete-valued vector representation $\hat{\jmb{y}} = E(\jmb{x})$ as well as  a decoder function $D(\cdot)$ that reconstructs the image from the latent code: $\hat{\jmb{x}} = D(\hat{\jmb{y}})$. Both the encoder and decoder are trained to optimize for a balance between minimizing the expected code length of the latent code and maximizing the quality of the reconstruction; this is otherwise known as the rate-distortion tradeoff
\begingroup\abovedisplayskip=3pt \belowdisplayskip=3pt
\begin{equation}
\begin{aligned} \label{eq:rate_distortion}
\ell(\jmb{x}, \hat{\jmb{x}})  + \beta R(\hat{\jmb{y}})
\end{aligned}
\end{equation} 
\endgroup
where $\ell$ is the reconstruction error between the input and the decoded output, typically measured by MSE (mean squared error) or a differentiable image quality metric such as MS-SSIM \cite{wang_msssim}; 
$R$ is the cost for encoding the latent representation to a bitstream measured by bitrate.  
The bitrate is oftentimes approximated in a differentiable manner by measuring the cross-entropy between the latent code distribution and a learned prior: 
\begingroup\abovedisplayskip=4pt \belowdisplayskip=4pt
\begin{equation}
R(\hat{\jmb{y}}) \approx \mathbb{E}_{\hat{\jmb{y}} \sim p_{\hat{\jmb{y}}}}[\log p(\hat{\jmb{y}}; \jmb{\theta})]
\end{equation}
\endgroup
Towards these goals, researchers have devised various ways to make the discrete binary encoding operations suitable for end-to-end learning, such as straight-through estimation \cite{toderici_varimgcomp, theis_imgcomp_ae}, soft quantization \cite{agustsson_softtohard, mentzer_condprobimg} and noise sampling \cite{balle_imgcomp_end2end, balle_varhyperprior}.  Moreover, sophisticated prior models have been designed for the quantized representation in order to minimize the cross-entropy with the code distribution. 
Different approaches to model the prior include autoregressive models \cite{minnen_jointpriors, mentzer_condprobimg, toderici_fullimgcomp}, hyperprior models \cite{balle_varhyperprior, minnen_jointpriors}, and factorized models \cite{theis_imgcomp_ae, balle_imgcomp_end2end, balle_varhyperprior}.

\subsection{Deep Video Compression} \label{sec:old_multiview}
Traditional video compression techniques exploit temporal redundancies by encoding independent frames (I-frames), then using motion compensation / residual coding to encode neighboring frames (P-frames, B-frames) \cite{hevc, wiegand_h264}. Recently, several deep-learning based video compression frameworks  \cite{wu_vidinterpolation, rippel_learnedvidcomp, han_vid_deepprob, liu_neuralvid} have been developed.  Wu~\etal \cite{wu_vidinterpolation} employs techniques based upon traditional video compression methods, while Rippel~\etal \cite{rippel_rtimgcomp} uses an end-to-end learning approach and achieves state-of-the-art results compared to traditional video codecs, including HEVC/H.265 and AVC/H.264. 

Video compression techniques may not necessarily translate directly to a stereo setting because they typically rely on temporal redundancies between a larger block of images for most bitrate savings. We show in our experiments that indeed motion/residual coding can struggle for stereo.

\subsection{Multi-view Compression}
There has been much prior work on designing and analyzing multi-view compression techniques, usually in a video compression setting \cite{flierl_disparitymvc, flierl_mvcoverview, merkle_mvc1, lukacs_multiviewpoint, martinian_viewsynth, kitahara_viewinterp}. In this setting, a multi-view video stream is reorganized as a matrix of pictures 
capturing temporal similarity between successive frames in one view and inter-view similarity between adjacent camera views. 
Numerous techniques \cite{merkle_mvc1, lukacs_multiviewpoint} use disparity compensated prediction to code each view given a reference view, similar to motion-compensated prediction in single-view video. The Multi-View Video Coding (MVC) extension developed for H.264/AVC uses this approach \cite{merkle_mvc1}. 
Other techniques involve using dense depth maps to synthesize a more precise view prediction for compression \cite{martinian_viewsynth, kitahara_viewinterp}.

Stereo specific compression techniques exist, and range from using a Markov random field (MRF) for disparity prediction to separate transforms for residual images \cite{siegel_stereostreams, frajka_resstereo, woo_stereomrf, aydinoglu_stereoproj, moellenhoff_codingres, tahara_stereoscopic}.
Also closely related is light field image compression, where  learning-based view synthesis techniques  are used to take advantage of the vast redundancy between the subaperture images \cite{jia_lf_gan, jiang_lf_view}.

In contrast, we use an end-to-end deep architecture for stereo image compression. Implicit depth estimation and compression is performed jointly in our model. 
\subsection{Stereo Matching} \label{sec:related_work}
Traditional stereo matching methods range from  local similarity estimation \cite{birchfield_ncc,ryan_ssd,hannah_sad}, particle propogation methods such as PatchMatch \cite{barnes_patchmatch}, to variational inference such as conditional random fields \cite{scharstein2007learning} and semi-global matching  (SGM) \cite{hirschmuller_sgbm}. 
There have been advances in deep learnable stereo matching, utilizing both supervised losses (training against ground truth disparities) \cite{zbontary_mccnn, luo_etal_cvpr16, chang_psmnet, kendall_gcnet} as well as unsupervised losses (using a photometric/consistency loss) \cite{zhong_selfsupstereo}. 
Stereo matching can be seen as a specific case of disparity-compensated prediction for the stereo image compression setting. 
Nonetheless, supervised learning-based stereo matching methods require ground-truth (GT) to train, and acquiring GT for stereo is difficult and expensive.
\section{Deep Stereo Image Compression}

In this paper we tackle the problem of compressing a pair of stereo images. 
Intuitively, if the overlapping field of view  between the stereo  pair 
is very high, then the bitrate of the combined latent code should be lower than the sum of the bitrates if we compressed the images separately; at the very least it cannot be higher. More formally, let us denote $\jmb{x}_1, \jmb{x}_2$ as a pair of rectified stereo images and let $H(\jmb{x}_1, \jmb{x}_2)$ be the entropy of the stereo pair. Given the fact that the content of the two images are highly correlated, the mutual information 
\[I(\jmb{x}_1, \jmb{x}_2) = H(\jmb{x}_1) + H(\jmb{x}_2) - H(\jmb{x}_1, \jmb{x}_2) \geq 0
\]

This observation motivates our general approach: we propose a single compression model that jointly compresses two stereo images. Towards this goal, we focus on extracting as much information as possible from one image in order to reduce the bitrate in the second, such that the total bitrate is lower than the result of independent single-image compression. Our approach is a two-stream deep encoder-decoder network as shown in Fig.~\ref{fig:stereo_complex_model}. Each image in a stereo pair is passed to a separate encoder/quantizer to get a discretized latent code; then, a decoder is utilized to reconstruct the image from the latent code. Compared to previous work, we have two major contributions: 1) we add \textit{multi-level, parametric skip functions} from the first image's feature maps to propagate information to the second image; 2) we utilize a \textit{conditional entropy model} to model the correlation between the latent codes of the two images. Next we will describe each component in details.

 \subsection{Encoding/Decoding and Quantization} \label{sec:feature_map}
Our encoder, decoder, and quantizer functions 
borrow their architectures from the single-image compression model in Ball\'e et al. \cite{balle_varhyperprior}.  As shown in Fig.~\ref{fig:stereo_complex_model}, each encoder is implemented with a series of 4 downsampling convolutions (by 2x) and  Generalized Divisive Normalization (GDN) layers \cite{balle_gdn}. Each decoder is implemented with a series of 4 upsampling deconvolutions (by 2x) and Inverse-GDN layers. Each quantizer $Q$ applies a rounding function to the floating-point output of the encoder $Q(E(\jmb{x})) = Q(\jmb{y}) = \hat{\jmb{y}}$ to output the discrete code representation.

\subsection{Parametric Skip Function} \label{sec:vol_disparity}
To reduce the joint bitrate across the stereo pair, we design a network module called a  \textit{parametric skip function} to propagate information from the first image's encoder/decoder to the second.
We conjecture that for a given stereo pair, there exists a correlation between the feature maps of the two images at all levels in the encoder and decoder. Moreover, if we estimate the disparity between each pair of feature maps, we can warp one feature to the other and improve the pixel-level alignment between the two feature maps; this in turn allows us to pass information from one feature map accurately to the corresponding spatial positions of the other. 

Specifically, in order to compute the feature map of the second image at level $t$, each skip function takes its previous layer's feature $\jmb{h}^{t-1}_2$, the previous layer feature from image one $\jmb{h}^{t-1}_1$ and the first image's code $\hat{\jmb{y}}_1$ as input. Each skip function module consists of four parts. First, a fully convolutional \textbf{global context} encoding module $f(\hat{\jmb{y}_1}; \jmb{w})$ encodes the first image's latent code to a feature descriptor $\jmb{d}_1$, to capture global context information of the first image, contained in its latent code. The global context feature is shared across all the different levels.  Secondly, a \textbf{stereo cost volume} module estimates a cost volume $\jmb{c}^{t-1}$ from the input of the first feature map, second feature map 
and the global context. 
The cost volume's size is $C\times H^{t-1} \times W^{t-1}$ where $C$ is the maximum disparity and $H^{t-1}$ and $W^{t-1}$ are the height/width of $\jmb{h}^{t-1}_1$. 
A softmax layer is applied to ensure the cost is normalized along the disparity dimension per pixel. Each value in the cost volume can be seen as a probability/confidence measure of the correct disparity at that coordinate. We then use this cost volume to \textbf{densely warp} the feature from the first image to the second. Particularly, for each pixel $i$ the new feature vector is computed through a weighted sum of feature vectors across all the disparity values in the disparity range:
\begin{equation}
\jmb{g}^{t-1}_{2, i} = \sum_{d=0}^{C} c_{d, i} \cdot \jmb{h}^{t-1}_{1,(i,d)}
\end{equation}
where $c_{d, i}$ represents the cost of disparity $d$ at pixel $i$. $(i,d)$ represents the pixel index that is $d$ pixels right of pixel $i$. 
The volumetric warping gives us a warped feature map $\jmb{g}^{t-1}_{2}$ which better aligns with the feature map of the second image; this can be also seen as an attention mechanism for each pixel $i$ into the first image's feature map within a disparity range. This design is conceptually similar to previous image synthesis work \cite{flynn_deepstereo, xie_deep3d}. Compared to regressing a single disparity map and warping with bilinear sampling \cite{jaderberg_spatialtransform}, our design allows more flexible connections between the target pixel and a range of pixels from the source image.  Finally, we use an \textbf{aggregation} function to predict the feature map as the final output of our parametric skip function:
\begin{equation}
\jmb{h}^{t}_2 = a(\jmb{g}^{t-1}_{2}, \jmb{h}^{t-1}_2)
\end{equation}
with $\jmb{g}^{t-1}_{2, i}$  the volumetric warped feature from the first iamge and $\jmb{h}^{t-1}_2$ the previous layer's feature from the second image. We refer to supplementary material (Sec. C)
for architecture details of context encoding, stereo cost volume and aggregation.
\begin{figure}
	\centering
	\includegraphics[width=.8\linewidth]{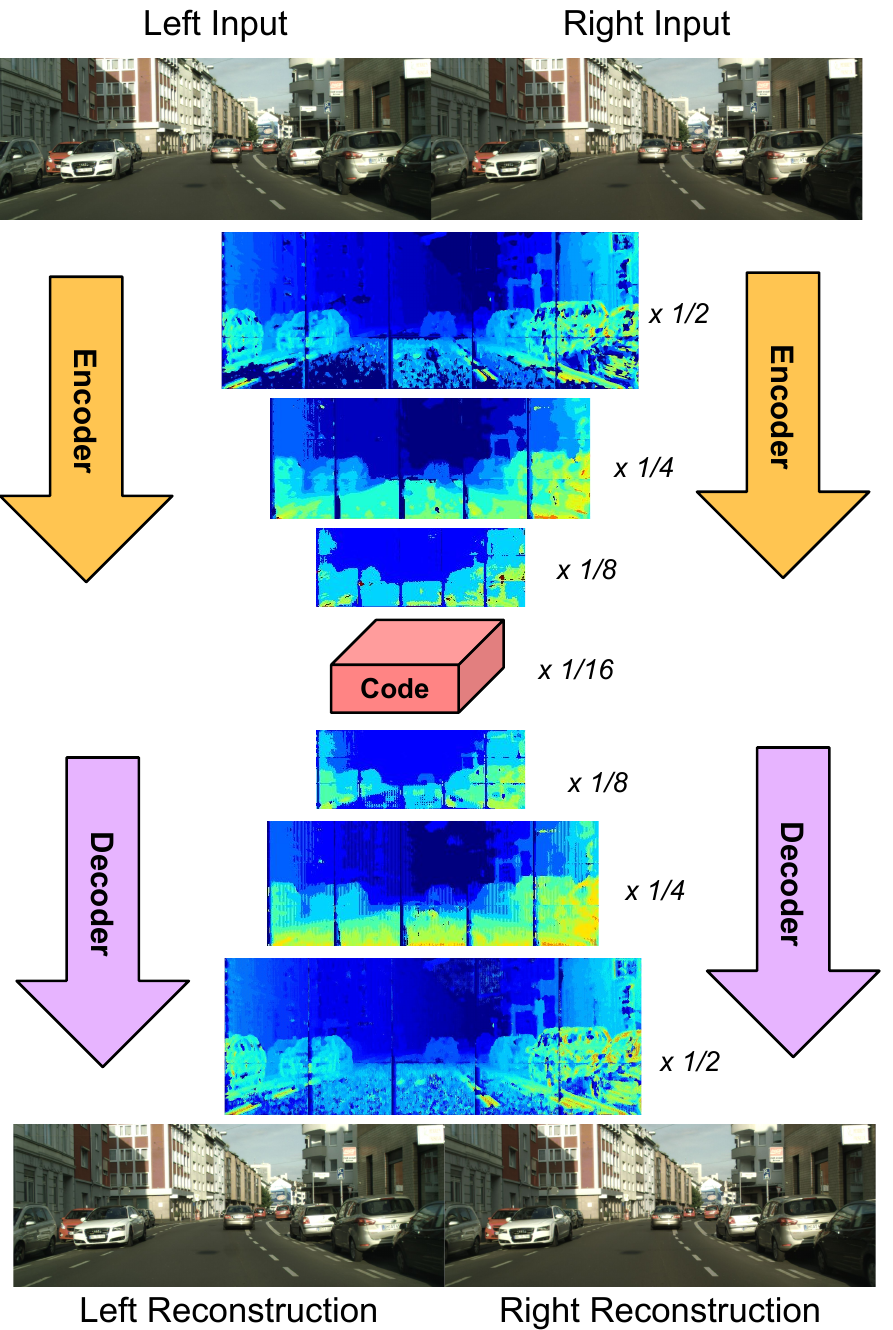}
	\caption{Visualization of the disparity volumes at each resolution level in the encoder/decoder, by taking the mode over the disparity dimension for each feature pixel. Tiling effects are inherently due to unsupervised training on crops.}
	\label{fig:disp_vis}
	\vspace{-4mm}
\end{figure}
\subsection{Conditional Entropy Model} \label{sec:entropy_model}

\begin{figure*}
	\centering
	\includegraphics[height=0.25\linewidth]{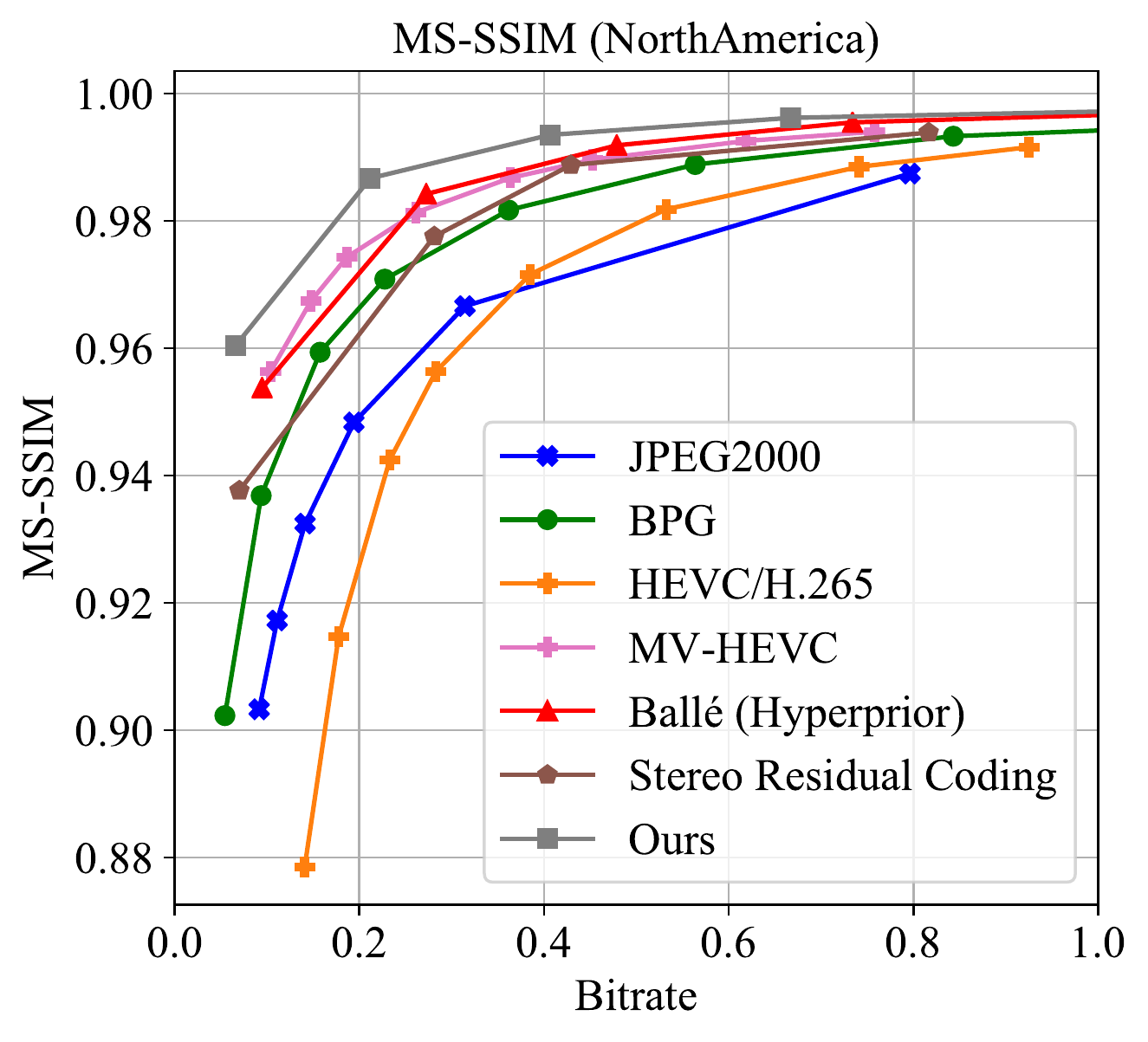}
        \includegraphics[height=0.25\linewidth]{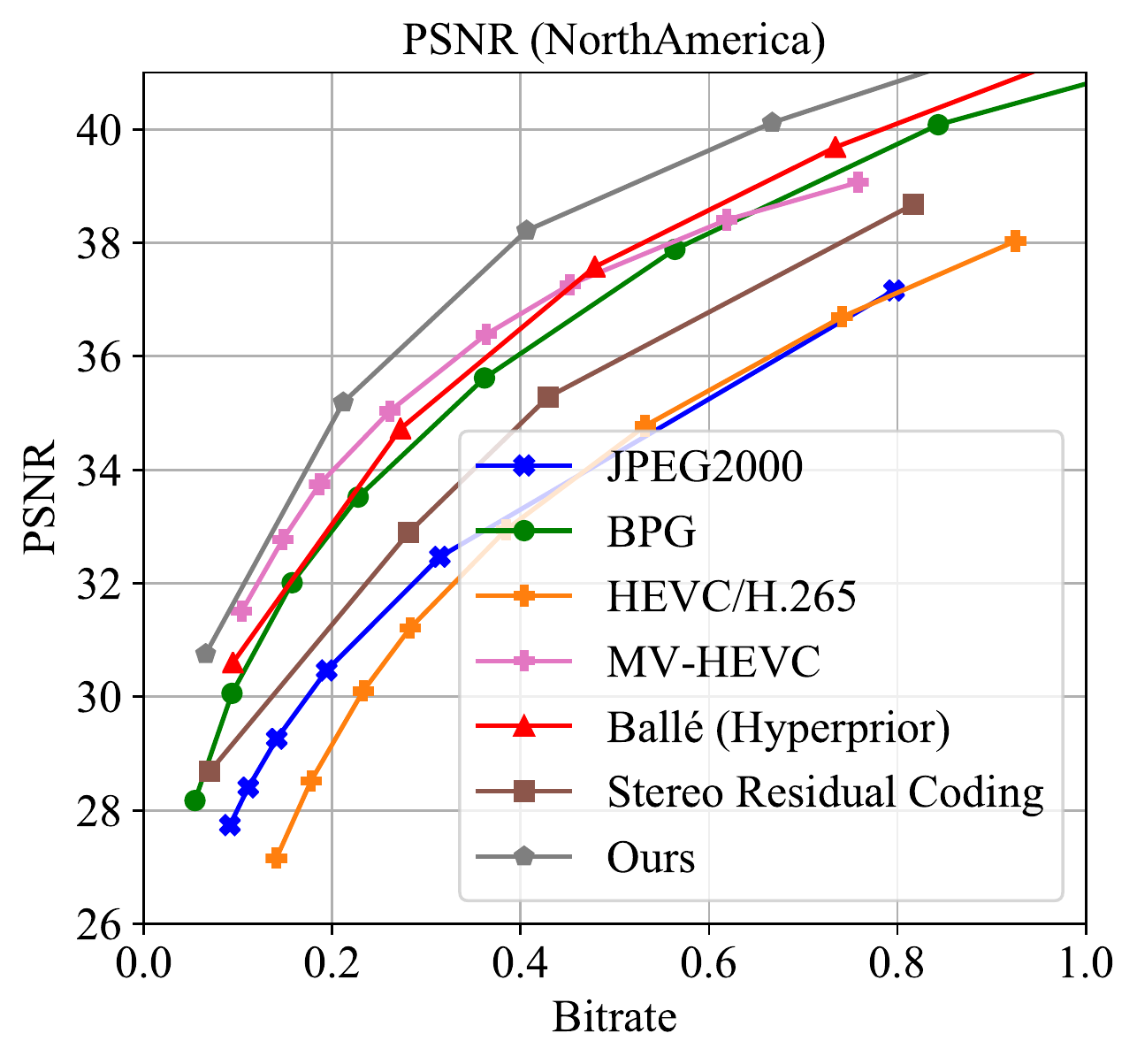}
	\includegraphics[height=0.25\linewidth]{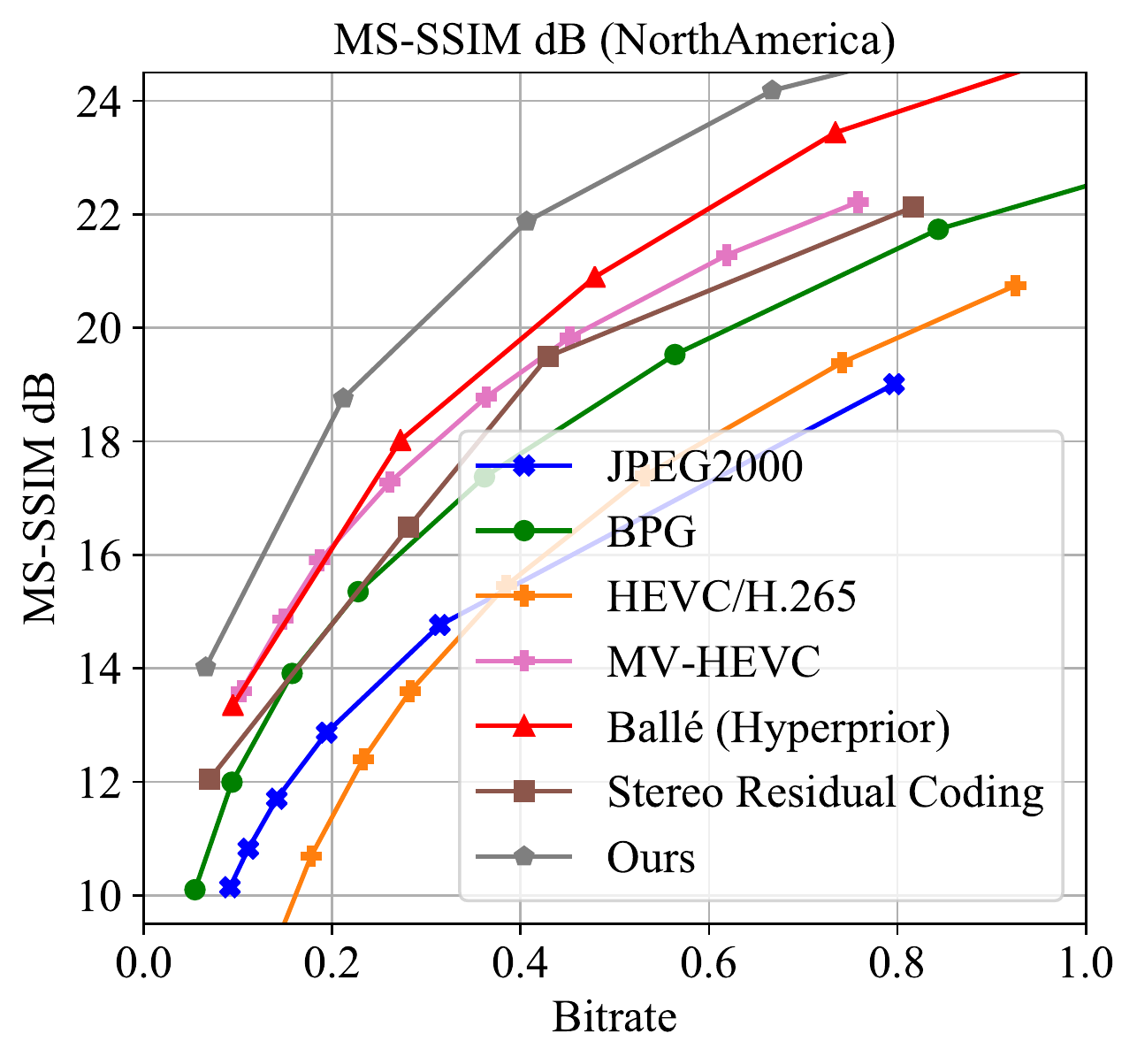} \\
	\includegraphics[height=0.25\linewidth]{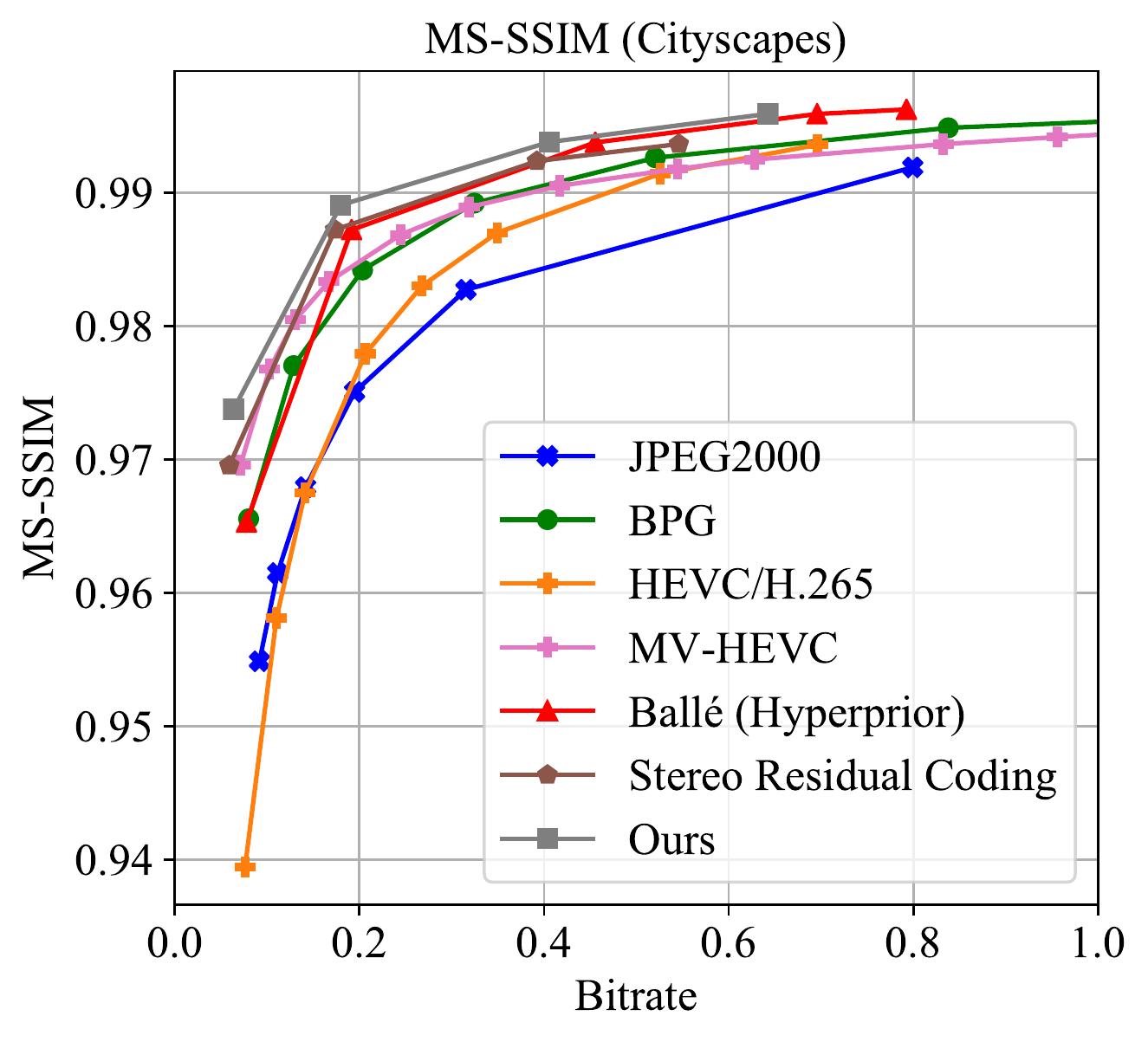}
        \includegraphics[height=0.25\linewidth]{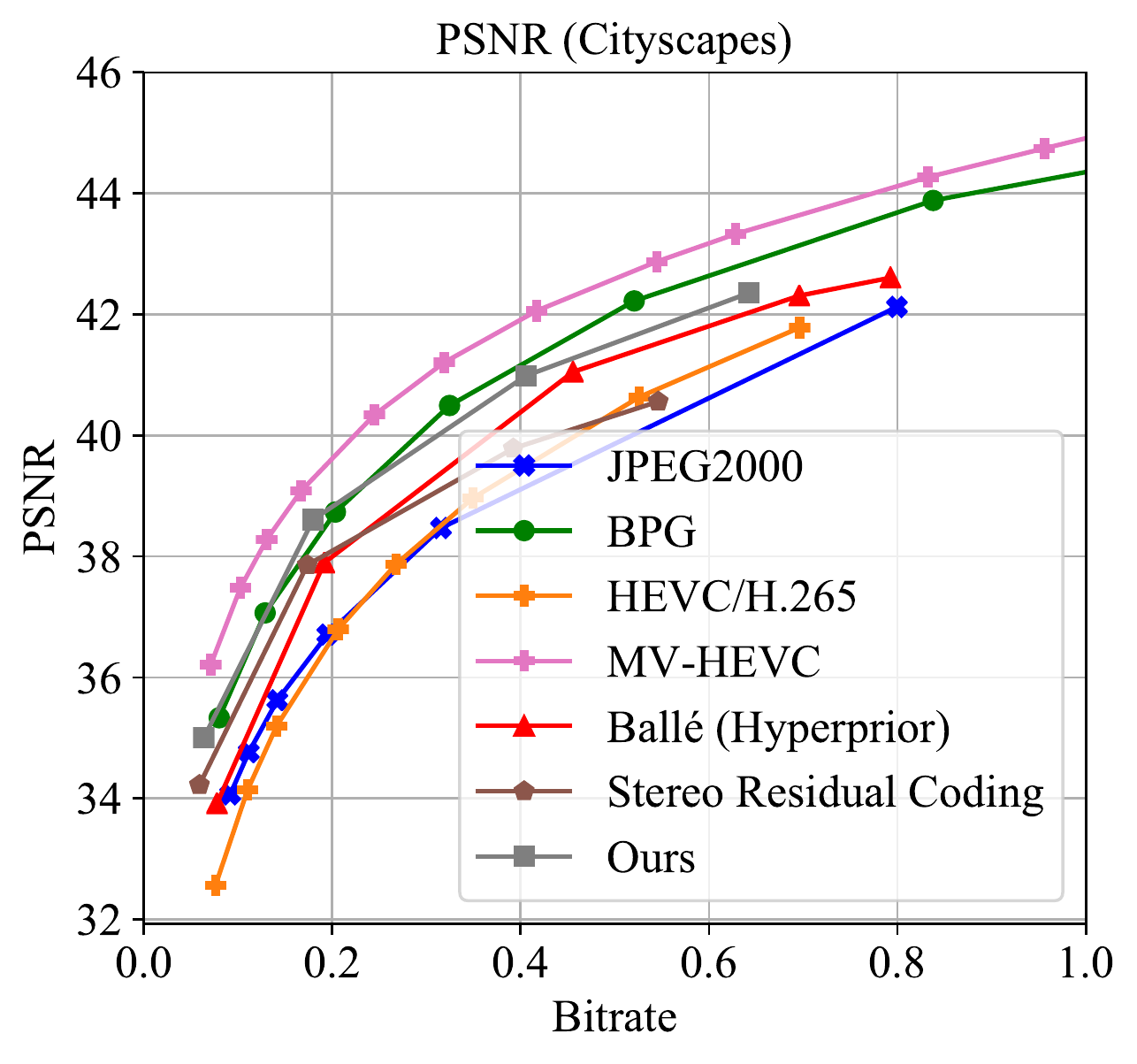}
	\includegraphics[height=0.25\linewidth]{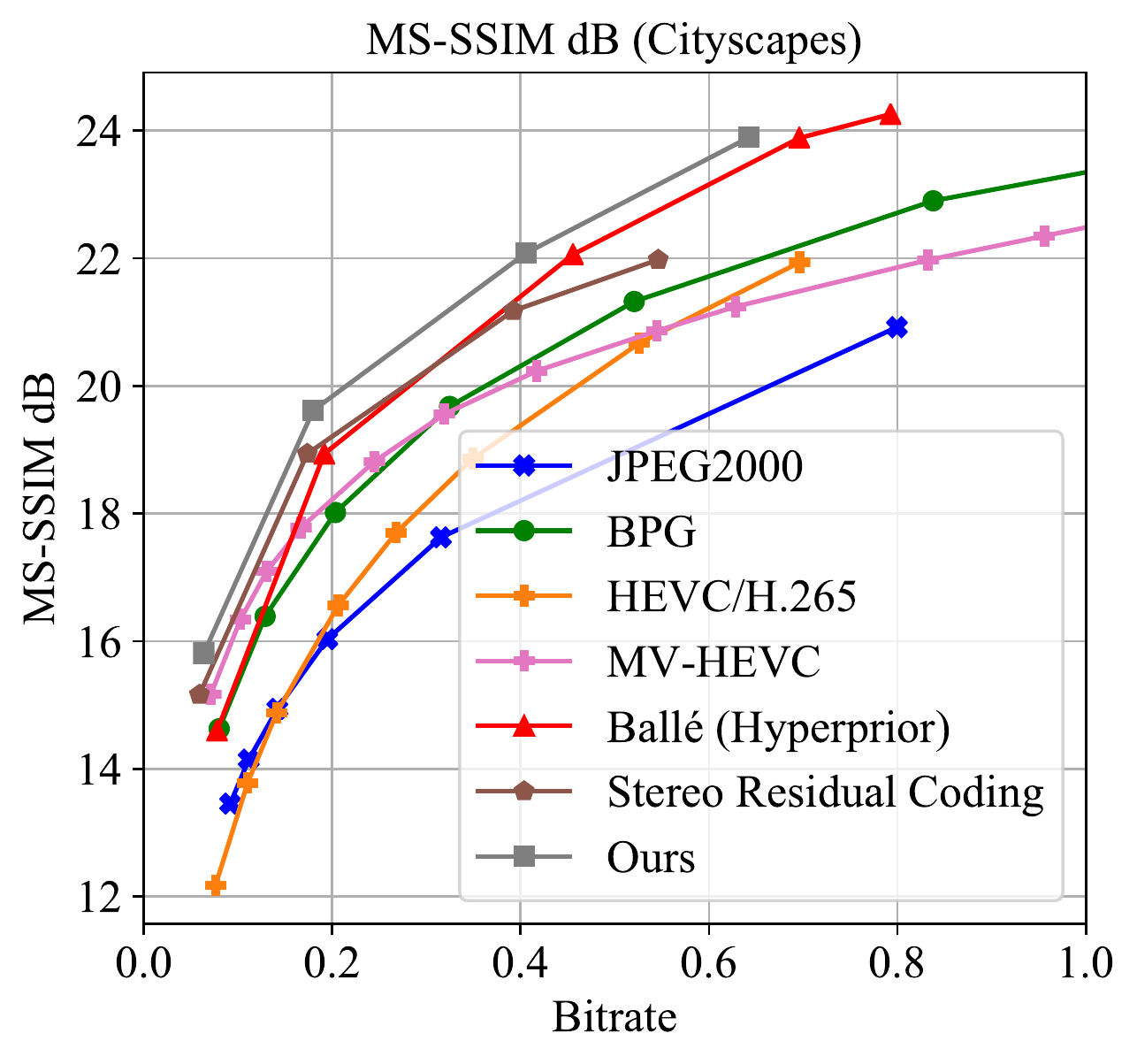}
		\vspace{-2mm}
	\caption{Plot of our stereo model against various compression baselines, for NorthAmerica and Cityscapes.}
	\label{fig:main_results_figure}
	\vspace{-4mm}
\end{figure*}

Accurately modeling the entropy of the quantized code is critical in navigating the rate-distortion trade-off, as entropy is a tight lower bound of bitrate \cite{shannon_lowerbound}. 
There exists a correlation between the latent codes of the two images, given the highly correlated image content. In order to exploit these relationships, we develop a joint entropy model with neural networks to estimate the joint distribution of the code. In order to maintain differentiability while mimicking the effect of rounding error during quantization, we consider a noisy version of $\jmb{y}$ as input: $\bar{\jmb{y}} = \jmb{y} + \jmb{\epsilon}$, where $\jmb{\epsilon} \sim \mathcal{U}(-0.5, 0.5)$. 

Our goal is to model the joint distribution $p(\bar{\jmb{y}}_1,\bar{\jmb{y}}_2 ; \jcb{\theta})$, which takes into account the dependence of $\bar{\jmb{y}}_2$ on $\bar{\jmb{y}}_1$. Inspired by \cite{balle_varhyperprior, minnen_jointpriors}, we additionally want to include \textit{side information as hyperpriors}, $\bar{\jmb{z}}_1, \bar{\jmb{z}}_2$, whose sole purpose is to reduce the entropy of $\bar{\jmb{y}}_1,\bar{\jmb{y}}_2$. Note that $\bar{\jmb{z}}_1, \bar{\jmb{z}}_2$ are derived from $\bar{\jmb{y}}_1,\bar{\jmb{y}}_2$ respectively during encoding and must also be counted in the bitstream. Thus, we factorize the joint probability of $\bar{\jmb{y}}_1,\bar{\jmb{y}}_2, \bar{\jmb{z}}_1, \bar{\jmb{z}}_2$ as follows: 
\begin{equation} \label{eq:stereo_entropy}
\begin{aligned}
p(\bar{\jmb{y}}_1,\bar{\jmb{y}}_2,\bar{\jmb{z}}_1,\bar{\jmb{z}}_2; \jcb{\theta}) = \\ p(\bar{\jmb{y} }_2| \bar{\jmb{y}}_1,\bar{\jmb{z}}_2; \jcb{\theta}_{\bar{\jmb{y}}_2}) p(\bar{\jmb{y}}_1 | \bar{\jmb{z}}_1 ; \jcb{\theta}_{\bar{\jmb{y}}_1})p(\bar{\jmb{z}}_2; \jcb{\theta}_{\bar{\jmb{z}}_2})p(\bar{\jmb{z}}_1; \jcb{\theta}_{\bar{\jmb{z}}_1})
\end{aligned}
\end{equation}
where $p(\bar{\jmb{y}}_1 | \bar{\jmb{z}}_1 ; \jcb{\theta}_{\bar{\jmb{y}}_1})$ denotes the probability 
of the first image code and $p(\bar{\jmb{y} }_2| \bar{\jmb{y}}_1,\bar{\jmb{z}}_2; \jcb{\theta}_{\bar{\jmb{y}}_2})$ denotes the probability
of the second image code, which is conditioned on the first image. $\jcb{\theta}_{\bar{\jmb{y}}_2},\jcb{\theta}_{\bar{\jmb{y}}_1},\jcb{\theta}_{\bar{\jmb{z}}_2},\jcb{\theta}_{\bar{\jmb{z}}_1}$ are the full set of parameters for each univariate distribution. 
All models are factorized into the product of each individual code's distribution under the full independence and conditional independence assumptions: 
\begin{align} \label{eq:ind_entropy}
p(\bar{\jmb{z}}_1; \jcb{\theta}_{\bar{\jmb{z}}_1}) &= \prod_i {p_{1,i}(\bar{z}_{1,i}; \jcb{\theta}_{\bar{\jmb{z}}_1})} \\
p(\bar{\jmb{z}}_2; \jcb{\theta}_{\bar{\jmb{z}}_2}) &=  \prod_i {p_{2,i}(\bar{z}_{2,i}; \jcb{\theta}_{\bar{\jmb{z}}_2})}\\
p(\bar{\jmb{y}}_1 | \bar{\jmb{z}}_1 ; \jcb{\theta}_{\bar{\jmb{y}}_1}) &= \prod_i {p_{1, i}(\bar{y}_{1,i} | \bar{\jmb{z}}_1; \jcb{\theta}_{\bar{\jmb{y}}_1})}  \\
 p(\bar{\jmb{y} }_2| \bar{\jmb{y}}_1,\bar{\jmb{z}}_2; \jcb{\theta}_{\bar{\jmb{y}}_2} ) &= \prod_i {p_{2, i}(\bar{y}_{2,i} | \bar{\jmb{y}}_1,\bar{\jmb{z}}_2; \jcb{\theta}_{\bar{\jmb{y}}_2})} 
\end{align}
Directly modeling a probability density function (PDF) with a deep parametric function may not be suitable for PDFs with discontinuous shapes, e.g., a uniform distribution between [-0.5, 0.5]. This restricts the power of the entropy model. Following \cite{balle_varhyperprior}, we overcome this issue by modeling probabilities as an area under the cumulative density function (CDF) as opposed to a point on the PDF. 
We first design our hyperprior models $p_{i}(\bar{z}_{i}; \jcb{\theta}_{\bar{\jmb{z}}})$ as follows:
\begin{equation} \label{eq:ind_entropy2}
\begin{aligned}
p_{i}(\bar{z}_{i}; \jcb{\theta}_{\bar{\jmb{z}}}) & = \left(q_{i} * u \right)(\bar{z}_{i}) 
\end{aligned}
\end{equation}
where $u(\tau) = 1 $ if $|\tau| < 0.5$ otherwise $u(\tau) = 0$, and  $*$ is the convolution between two functions. Thus we have:
\begin{equation} \label{eq:ind_entropy2}
\begin{aligned}
p_{i}(\bar{z}_{i}; \jcb{\theta}_{\bar{\jmb{z}}})  &  =\int_{-\infty}^{\infty} q_{i}(\tau; \jcb{\theta}_{\bar{\jmb{z}}}) u(\bar{z}_{i} - \tau) d\tau
\\&  =\int_{\bar{z}_{i}-0.5}^{\bar{z}_{i}+0.5} q_{i}(\tau; \jcb{\theta}_{\bar{\jmb{z}}}) d\tau \\
 & = c_{i}(\bar{z}_{i}+ 0.5;\jcb{\theta}_{\bar{\jmb{z}}}) - c_{i}(\bar{z}_{i} - 0.5; \jcb{\theta}_{\bar{\jmb{z}}})
\end{aligned}
\end{equation}
where $c_{i}(\bar{z}_{i}; \jcb{\theta}_{\bar{\jmb{z}}})$ is the cumulative density function (CDF) of some underlying PDF $q$.
This intuitively means that we can define $p_{i}(\bar{z}_{i}; \jcb{\theta}_{\bar{\jmb{z}}})$ as an area under the CDF rather than directly as the PDF, and we can use a neural network to directly model $c_{i}(\bar{z}_{i}; \jcb{\theta}_{\bar{\jmb{z}}})$. This approach has better capacity to model steep edge PDFs, since even for steep edged PDF, the CDF is still continuous. 

We follow a similar approach to model the conditional factorized probabilities for $\bar{\jmb{y}}_1, \bar{\jmb{y}}_2$ - we first highlight the model for $\bar{\jmb{y}}_2$: 
\begin{equation}
p_{2, i}(\bar{y}_{2,i} | \bar{\jmb{y}}_1,\bar{\jmb{z}}_2; \jcb{\theta}_{\bar{\jmb{y}}_2}) =  \left(q_{2, i} * u \right)(\bar{y}_{2,i}) 
\end{equation}
 Unlike the hyperprior models, we model each individual pixel PDF $q_{2, i}$ as a Gaussian mixture model (GMM):
 \begin{equation}
 q_{2, i}(\bar{\jmb{y}}_1,\bar{\jmb{z}}_2) = \sum_k{w_{ik} \mathcal{N}(\mu_{ik},\sigma^2_{ik})})
 \end{equation}
where $w_{ik}, \mu_{ik}, \sigma^2_{ik}$ are the distribution parameters depending on $\bar{\jmb{y}}_1,\bar{\jmb{z}}_2$ and $\jcb{\theta}_{\bar{\jmb{y}}_2}$. We also rewrite the convolution as the difference between CDFs as in Eq.~(\ref{eq:ind_entropy2}). The CDF of a GMM is generally computed numerically in most computational frameworks, while the derivative is analytical. Thus we just need to learn a function that predicts parameters  $w_{ik}, \mu_{ik}, \sigma^2_{ik}$ given $\bar{\jmb{y}}_1,\bar{\jmb{z}}_2$ with $\jcb{\theta}_{\bar{\jmb{y}}_2}$ as learnable parameters, instead of modeling the CDF value directly as in the hyperprior entropy model. We found that a mixture model increased performance slightly thanks to its stronger capacity compared to a single Gaussian. 
Finally, the model for $\bar{\jmb{y}}_1$ follows the same GMM formulation; however given that $\bar{\jmb{y}}_1$ is decoded first, we can only provide $\bar{\jmb{z}}_1$ as input, not $\bar{\jmb{y}}_2$:
\begin{equation}
p_{1, i}(\bar{y}_{1,i} | \bar{\jmb{z}}_1; \jcb{\theta}_{\bar{\jmb{y}}_1}) =  \left(q_{1, i} * u \right)(\bar{y}_{1,i}) 
\end{equation}
Architecture details for our hyper-encoder - deriving $\bar{\jmb{z}}_1, \bar{\jmb{z}}_2$ from $\bar{\jmb{y}}_1, \bar{\jmb{y}}_2$ - as well as for each entropy model can be found in supplementary material (Sec C.3).

\subsection{Learning}
Our model is trained end-to-end to minimize the following objective function: 
\begin{equation}
\begin{aligned} \label{eq:total_loss}
\ell + \beta R = \mathbb{E}_{\jmb{x}_1, \jmb{x}_2 \sim p_{\jmb{x}}}[ \underbrace{{||\jmb{x}_1 - \hat{\jmb{x}}_1||}^2_2}_{\text{Distortion (Img. 1)}} + \underbrace{{||\jmb{x}_2 - \hat{\jmb{x}}_2||}^2_2}_{\text{Distortion (Img. 2)}} ] + \\ 
\beta \mathbb{E}_{\jmb{x}_1, \jmb{x}_2 \sim p_{\jmb{x}}}[\underbrace{-\log_2 p(\bar{\jmb{y}}_1,\bar{\jmb{z}}_1; \jcb{\theta})}_{\text{Rate (Code 1)}}  \underbrace{- \log_2 p(\bar{\jmb{y}}_2,\bar{\jmb{z}}_2 | \bar{\jmb{y}}_1; \jcb{\theta})}_{\text{Rate (Code 2)}}]
\end{aligned}
\end{equation} 
where the first term encodes reconstruction quality of both images and the second term is the bitrate term with the rate predicted by the entropy model. 
Moreover, we can enforce a target bitrate for a given model by modifying the rate function to be: \begin{equation}
\begin{aligned} \label{eq:target_entropy}
R = \max(\mathbb{E}_{\jmb{x}_1, \jmb{x}_2 \sim p_{\jmb{x}}}[-\log_2 p(\bar{\jmb{y}}_1, \bar{\jmb{y}}_2,\bar{\jmb{z}}_1,\bar{\jmb{z}}_2; \jcb{\theta})], H_t)
\end{aligned}
\end{equation} 
where  $H_t$ is our desired target entropy calculated from the target bitrate. 

\newcommand{\jcaption}[1]{\put(0,57){\tiny \colorbox{gray}{\color{white} #1}}}
\newcommand{\jcaptionb}[1]{\put(0,56){\tiny \colorbox{gray}{\color{white} #1}}}
\newcommand{\torboxl}[0]{\put(33,16){\color{green}\framebox(34,28){}}
	\put(0,0){\setlength{\fboxsep}{0pt}\fbox{\includegraphics[viewport=120 120 240 240, clip, width=34pt]{figures/qual_eval/crops_tor4d_1/test_in1_tor4d_2x.png}}}
}
\newcommand{\torboxr}[0]{\put(38,16){\color{green}\framebox(30,28){}}}
\newcommand{\cityboxl}[0]{\put(42,10){\color{white} \framebox(25,32){}}}
\newcommand{\cityboxr}[0]{\put(44,10){\color{white} \framebox(28,32){}}}
\newcommand{\cityboxlb}[0]{\put(14,50){\color{white} \framebox(10,10){}}}
\newcommand{\cityboxrb}[0]{\put(16,50){\color{white} \framebox(10,10){}}}

\newcommand{\cityfigaa}[2]{
	\begin{overpic}[width=\imw]{#1}
		\put(18,35){ \color{green} \framebox(18, 15){}} 
		\put(49,0){\setlength{\fboxsep}{0pt}\color{orange}\fbox{\includegraphics[viewport=89 169 178 242, clip, width=50pt]{#1}}}
		\jcaptionb{#2}
\end{overpic}}
\newcommand{\cityfigab}[2]{
	\begin{overpic}[width=\imw]{#1}
		\put(20,35){ \color{green} \framebox(18, 15){}} 
		\put(49,0){\setlength{\fboxsep}{0pt}\color{orange}\fbox{\includegraphics[viewport=89 169 178 242, clip, width=50pt]{#1}}}
		\jcaptionb{#2}
\end{overpic}}
\newcommand{\torfigaa}[2]{
	\begin{overpic}[width=\imw]{#1}

		\put(77,15){ \color{green} \framebox(19, 16){}} 
		\put(0,0){\setlength{\fboxsep}{0pt}\color{orange}\fbox{\includegraphics[viewport=377 74 470 151, clip, width=56pt]{#1}}}
		\jcaptionb{#2}
\end{overpic}}
\newcommand{\torfigab}[2]{
	\begin{overpic}[width=\imw]{#1}

		\put(79,15){ \color{green} \framebox(19, 16){}} 
		\put(0,0){\setlength{\fboxsep}{0pt}\color{orange}\fbox{\includegraphics[viewport=387 74 480 151, clip, width=56pt]{#1}}}
		\jcaptionb{#2}
\end{overpic}}

\begin{figure*}
	\centering
	\def\imw{0.2\textwidth}
	\setlength{\tabcolsep}{1pt}
	\begin{tabular}{ccccc}
		\raisebox{4pt}{\footnotesize Input} & 
		\raisebox{4pt}{\footnotesize JPEG2000} &
		\raisebox{4pt}{\footnotesize BPG} & 
		\raisebox{4pt}{\footnotesize Ball\'e} &
		\raisebox{4pt}{\footnotesize Ours}
		\\ [-1pt]
		
		\cityfigaa{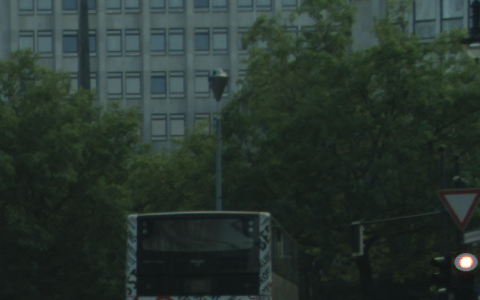}{\textbf{Cityscapes (Cam 1)}}
		&
		\cityfigaa{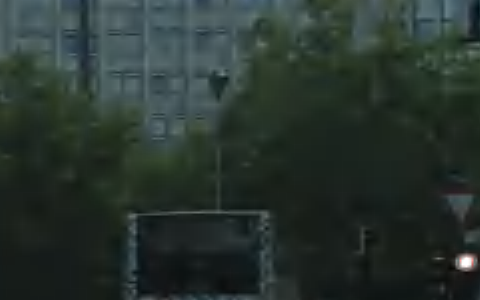}{Bitrate: 0.0648, PSNR: 33.11}
		&
		\cityfigaa{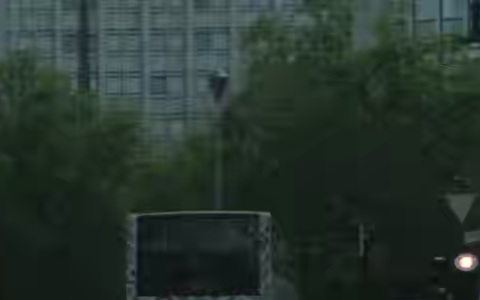}{Bitrate: 0.0651, PSNR: 34.59}
		&
		\cityfigaa{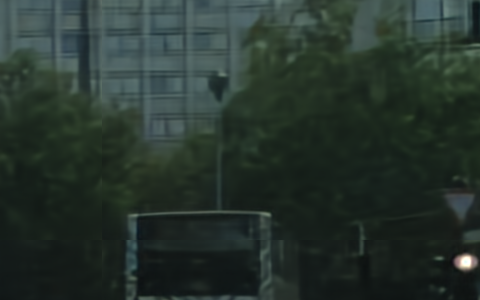}{Bitrate: 0.0770, PSNR: 34.62}
		& 
		\cityfigaa{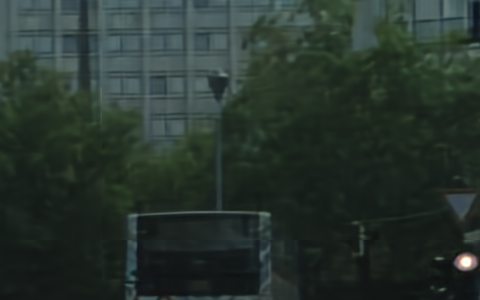}{\textbf{Bitrate: 0.0982, PSNR: 36.23}}
		\\
		\cityfigab{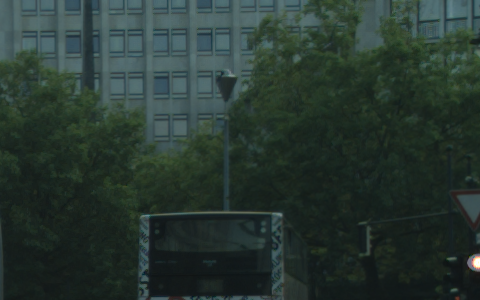}{\textbf{Cityscapes (Cam 2)}}
		&
		\cityfigab{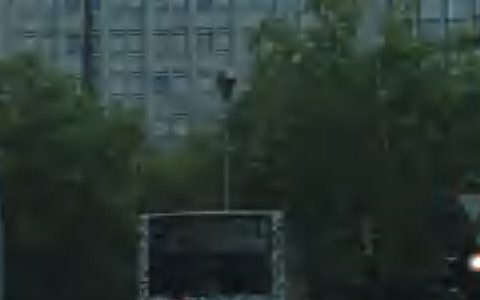}{Bitrate: 0.0643, PSNR: 32.71}
		&
		\cityfigab{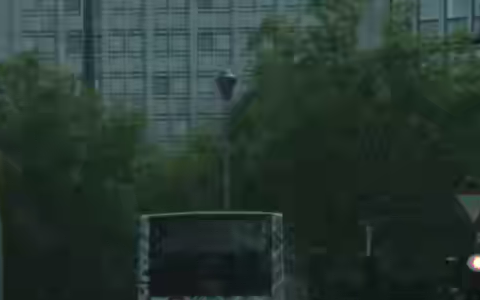}{Bitrate: 0.0649, PSNR: 34.38}
		&
		\cityfigab{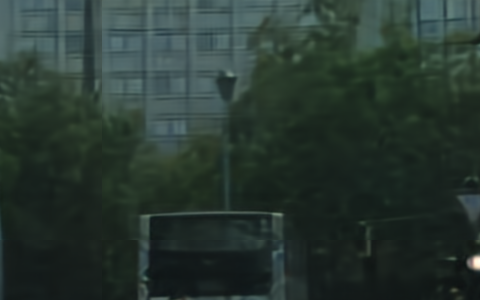}{Bitrate: 0.0792, PSNR: 34.16}
		& 
		\cityfigab{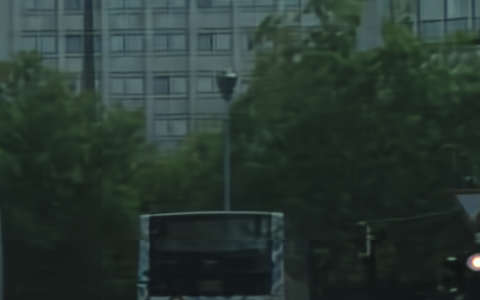}{\textbf{Bitrate: 0.0295, PSNR: 35.05}}
		\\ 
		\torfigaa{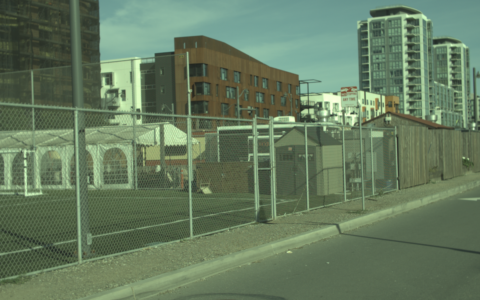}{\textbf{NorthAmerica (Cam 1)}}
		&
		\torfigaa{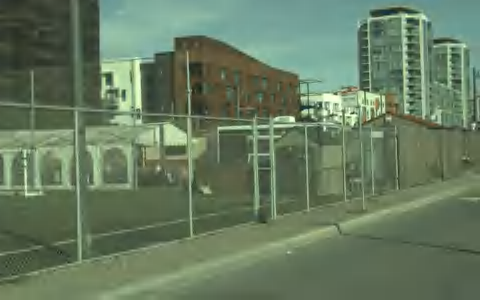}{Bitrate: 0.2825, PSNR: 30.64}
		&
		\torfigaa{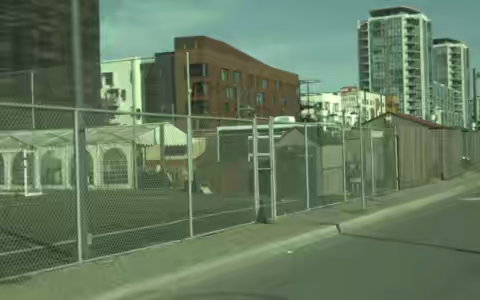}{Bitrate: 0.278, PSNR: 32.24}
		&
		\torfigaa{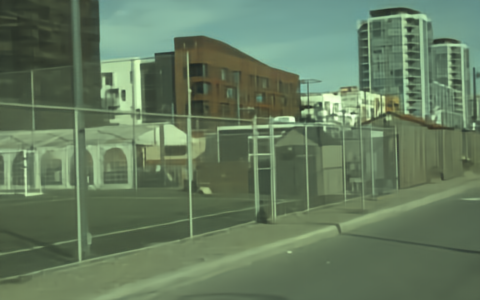}{Bitrate: 0.319, PSNR: 33.17}
		& 
		\torfigaa{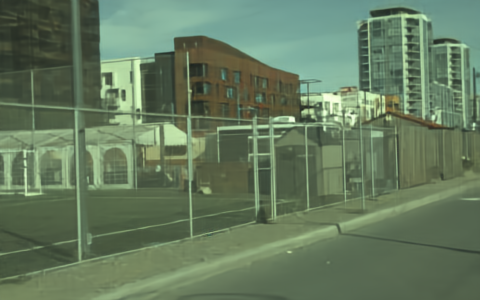}{\textbf{Bitrate: 0.321, PSNR: 33.71}}
		\\
		\torfigab{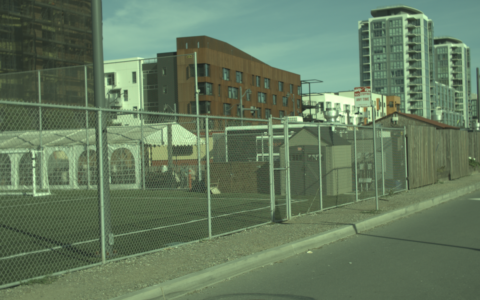}{\textbf{NorthAmerica (Cam 2)}}
		&
		\torfigab{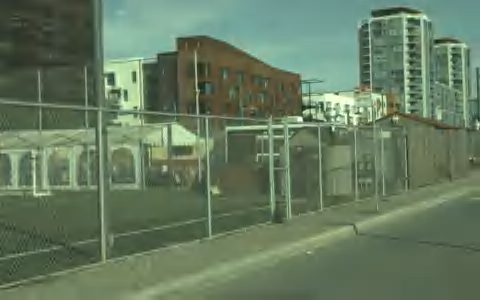}{Bitrate: 0.2838, PSNR: 30.43}
		&
		\torfigab{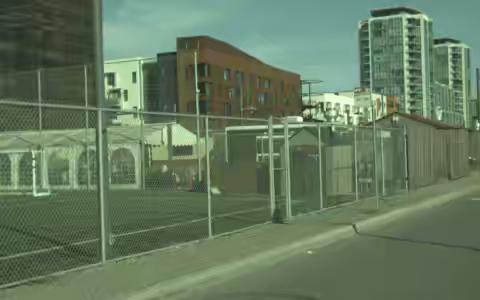}{Bitrate: 0.281, PSNR: 32.14}
		&
		\torfigab{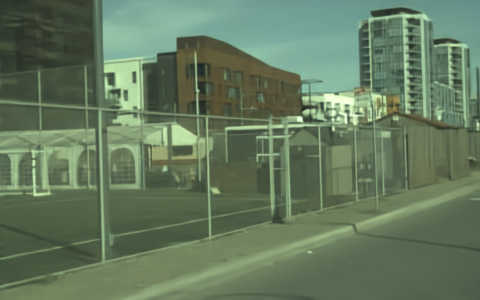}{Bitrate: 0.322, PSNR: 33.00}
		& 
		\torfigab{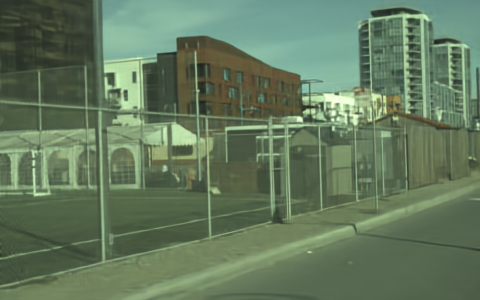}{\textbf{Bitrate: 0.200, PSNR: 33.24}}

	\end{tabular}
	\vspace{-2mm}
	\caption{Qualitative results of our method vs. various single-image baselines. Metrics are specified for each entry.}
	\label{fig:big_qual_results}
	\vspace{-2mm}
\end{figure*}

\section{Experiments}

We present a quantitative and qualitative evaluation of our approach and various baselines on two different datasets.
We now provide more details about the datasets and metrics we employ. 

\subsection{Datasets, Metrics and Baselines}
\vspace{-2pt}
\paragraph{NorthAmerica:}
We created a dataset captured by driving  self driving vehicles  in two different cities in North America.
This dataset consists of 100k pairs of lossless rectified stereo images. We use 2.5k pairs for validation, 18k for final testing, and the remaining for training. 
The images in this dataset are $480 \times 300$. 
\vspace{-4pt}
\paragraph{Cityscapes:} 
We also train on Cityscapes raw sequences \cite{cityscapes}, consisting of ~89k training pairs and 45k test pairs. For each $2048 \times 1024$ image, as a preprocessing step we crop 64 pixels from the top and 128 pixels from the left to remove rectification artifacts. We also crop out the bottom 240 pixels to remove the ego-vehicle in order to focus on the scene imagery. 
\vspace{-4pt}
\paragraph{Metrics:}
 We report results on both peak signal-to-noise ratio (PSNR) and multi-scale structural similarity (MS-SSIM) \cite{wang_msssim} as a function of bitrate. Both MS-SSIM and PSNR are commonly used perceptual quality metrics, and we measure both to test the robustness of our model. PSNR is defined as $-10\log_{10}(\text{MSE})$, where MSE is mean-squared error, and better measures the absolute error in a compressed image. On the other hand, MS-SSIM better measures how well overall structural information is preserved.  For the MS-SSIM curve, we report both at original scale as well as log-scale, namely "MS-SSIM (dB)" from \cite{balle_varhyperprior}, defined as $-10\log_{10}(1-\text{MS-SSIM})$. 
 
 \paragraph{Baselines:} 
Our completing algorithms include the single-image hyperprior Ball\'e model \cite{balle_varhyperprior} as well as popular image codecs - BPG and  JPEG2000.  We also try adapting traditional video-compression techniques as additional baselines. 
Specifically, we run a codec based on the HEVC/H.265 standard \cite{hevc} over the stereo pair; we also try a Multi-View Coding (MVC) extension of HEVC \cite{mv_hevc_github}.
Another approach is to try a deep learning method to encode the first image, disparity map and disparity-warped residual image jointly (referred to as ``stereo residual coding''): we compress the first image using the Ball\'e hyperprior model. Then we use SGM on the stereo pair to generate disparity estimates, and compress them using a second Ball\'e model. Finally, we compress the disparity-compensated residual image using a third Ball\'e model. 

\subsection{Implementation Details} 

We create multiple stereo compression models, each set to a different desired target bitrate. We set $\beta$, the weight on the entropy loss, to a value within 0.5 to 0.001 for the lower to higher bitrate models respectively.
For each model at a given target bitrate, we initialize the layers of both encoders and decoders with those from a pre-trained single-image based Ball\'e model \cite{balle_varhyperprior} at the corresponding bitrate. This speeds up our training and convergence time significantly.
We use a learning rate of $2 \cdot 10^{-4}$ for all models and optimize our parameters with Adam. We train with a total batch size of 4, across 4 GPUs. For NorthAmerica, we train on the full $480 \times 300$ image and set $C$ (the maximum disparity) to 32, while for Cityscapes, we train on $384 \times 240$ crops and set $C=64$.  

\begin{figure*}
	\centering
	\includegraphics[height=0.25\linewidth]{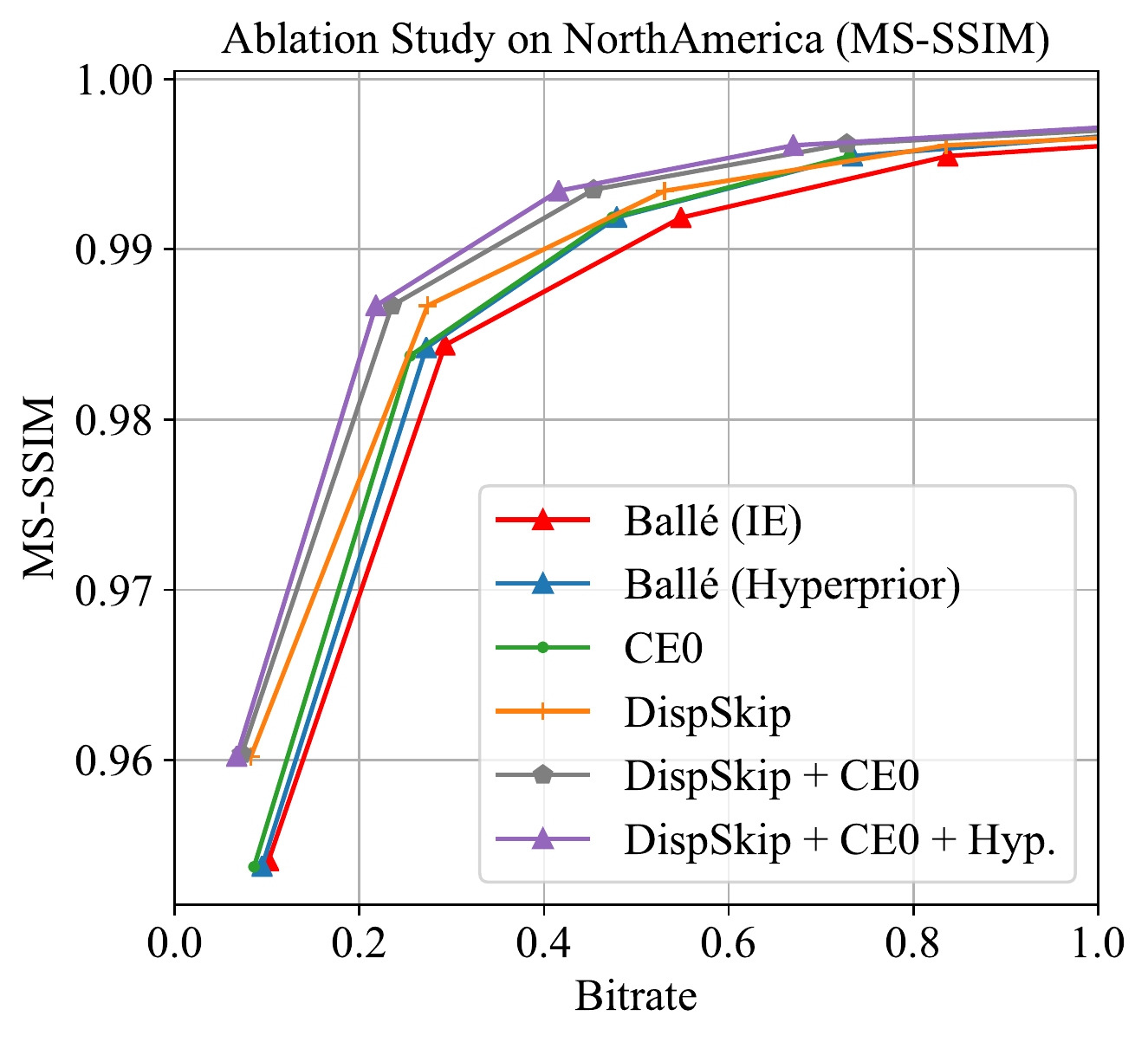}
        \includegraphics[height=0.25\linewidth]{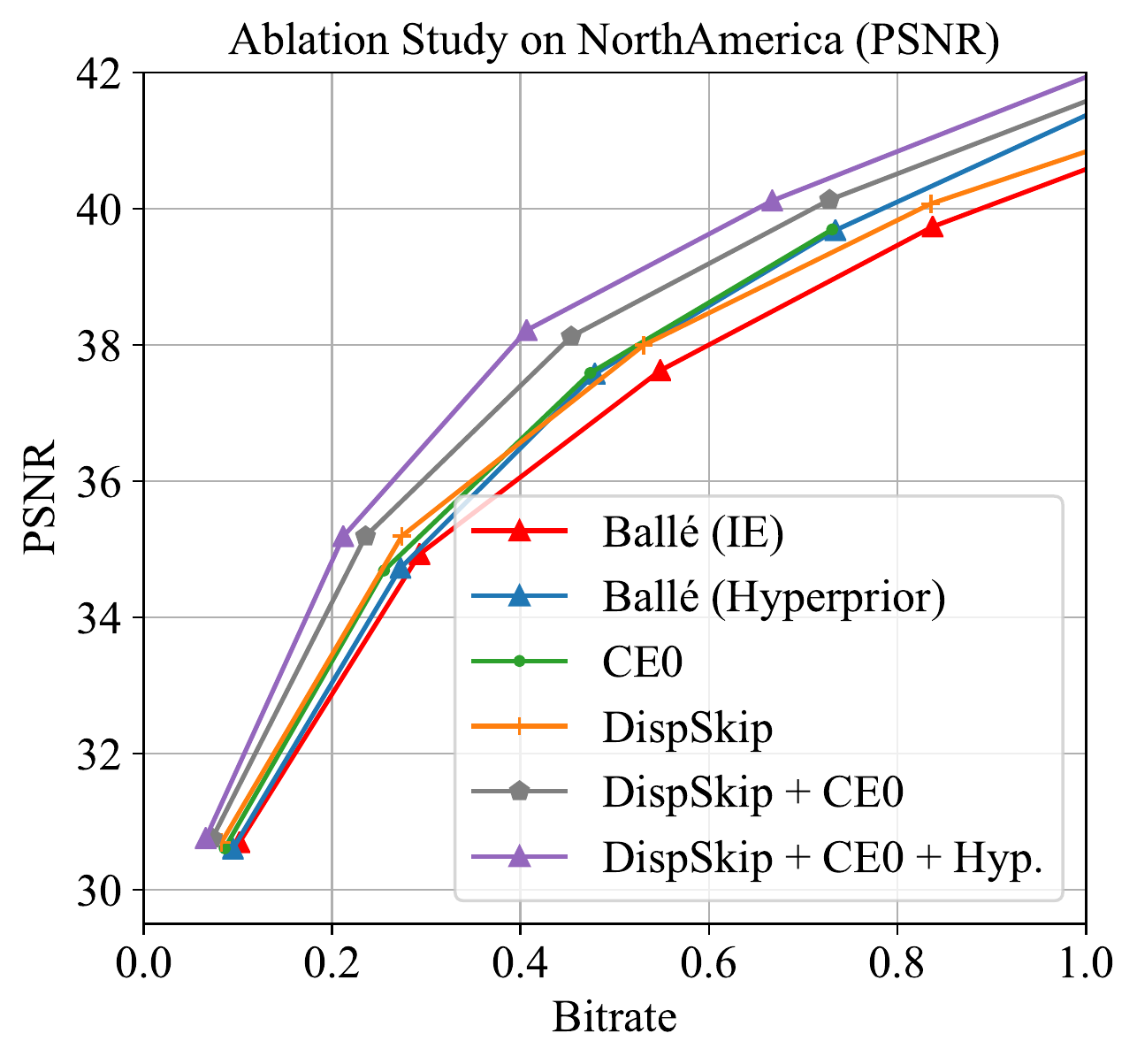}
	\includegraphics[height=0.25\linewidth]{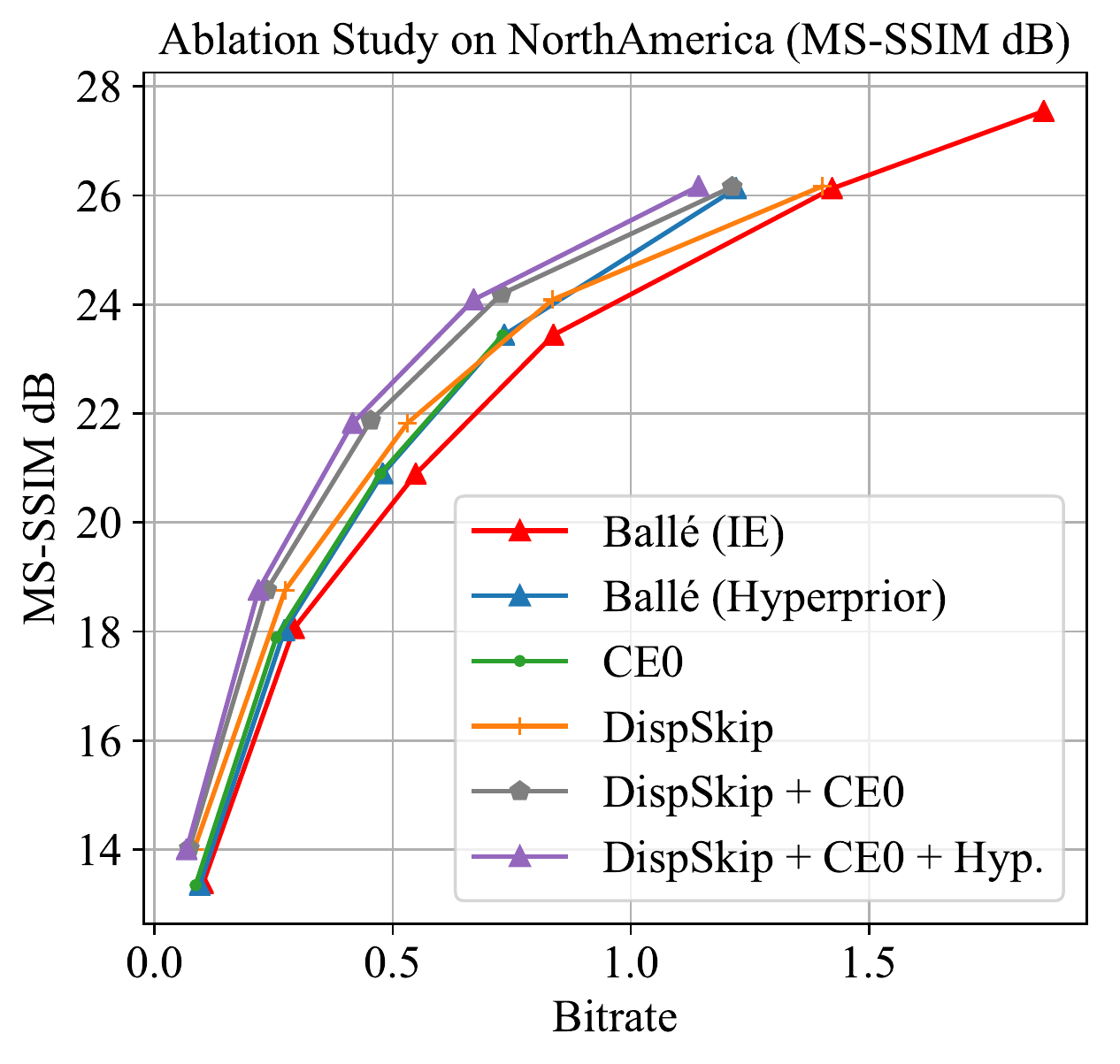} \\
		\includegraphics[height=0.25\linewidth]{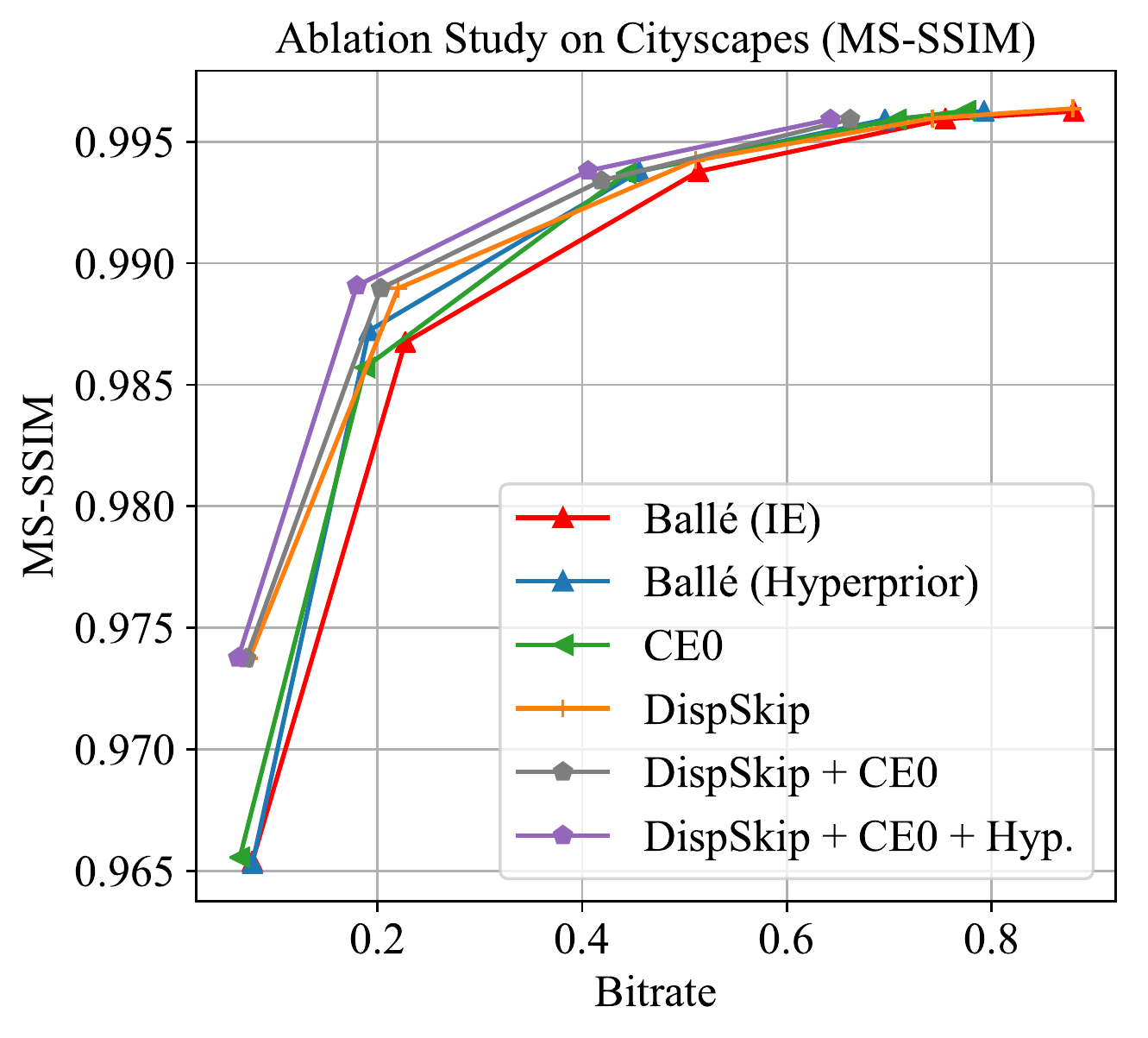}
        \includegraphics[height=0.25\linewidth]{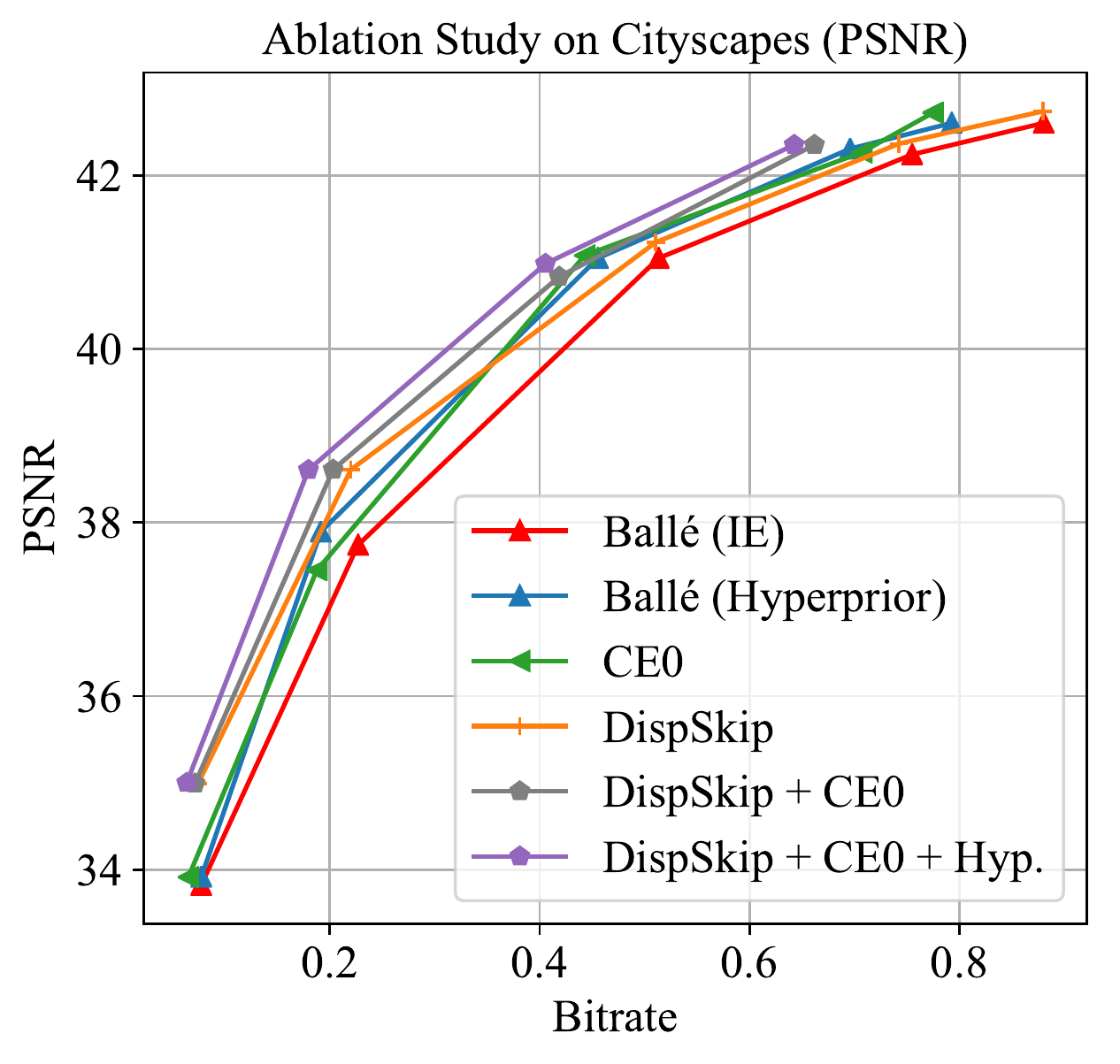}
	\includegraphics[height=0.25\linewidth]{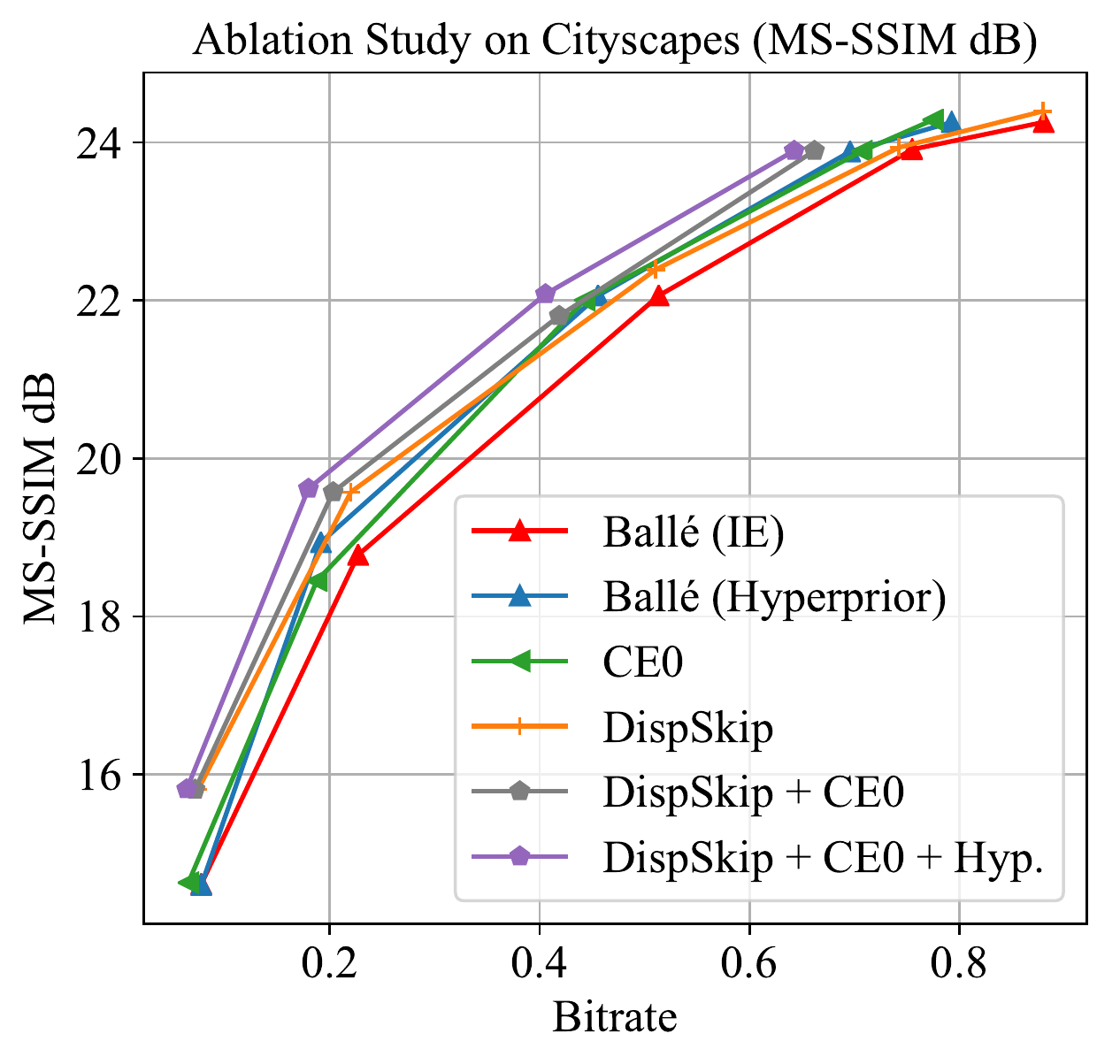}
		\vspace{-2mm}
	\caption{Ablation study. For both datasets, we analyze the independent and combined effects of our skip functions (DispSkip), conditional entropy without hyperprior (CE0), and hyperprior.}
	\label{fig:abl_results_figure}
	\vspace{-2mm}
\end{figure*}

\subsection{Experimental Results}
\vspace{-2pt}
\paragraph{Comparison to Baselines:}
On NorthAmerica, our stereo model consisting of our skip functions and conditional entropy outperforms all of these compression baselines, as shown in Fig. \ref{fig:main_results_figure}. 
Note that the reported "bitrate" for our stereo models represents the \textit{joint bitrate}, or the average of the bitrate for both images. Assuming our stereo pair is compressed as well under our model as under a single-image setting, this implies our \textit{residual bitrate} savings, the bitrate savings for the second image, to be double our joint bitrate savings. For NorthAmerica, at bitrate 0.0946 for the Ball\'e model, our model achieves an average $29.0\%$ reduction in the joint bitrate with higher MS-SSIM/PSNR, implying a $58.0\%$ reduction in the residual. At a higher Ball\'e bitrate 0.734, our model achieves a $17.8\%$ reduction in the residual with higher MS-SSIM/PSNR. 

For Cityscapes, our model outperforms all competing baselines in MS-SSIM; 
it outperforms all baselines except BPG/MV-HEVC in PSNR. 
At equal PSNR/MS-SSIM, residual savings against Ball\'e range from 30\% in the lower bitrates to 10\% at higher bitrates. 

Our deep residual coding baseline performs worse than single-image compression at all bitrates for NorthAmerica, and at higher bitrates for Cityscapes. Even though it numerically approaches the performance of our model at lower bitrates, there exist significant boundary/warping artifacts, which we demonstrate in supplementary material. The underperformance of our baseline is consistent with the findings of \cite{wu_vidinterpolation} and \cite{frajka_resstereo}, the latter of whom noted that residual images exhibit different correlation properties than the full images and may need to be modeled differently.

\vspace{-5pt}
\paragraph{Ablation Study:} We perform an ablation study in NorthAmerica and Cityscapes in Fig. \ref{fig:abl_results_figure} to isolate the impact of each component in our stereo model: the skip function (DispSkip), conditional entropy (separated from the hyperprior), and hyperprior. We start out with two fully factorized, independent single-image compression models (Ball\'e \cite{balle_varhyperprior}, denoted as IE), with no skip functions. We then analyze the impact of adding DispSkip on the IE model to isolate the effect of DispSkip from our full conditional entropy model. Then, we define a stripped down conditional entropy model (denoted as CE0) that removes all hyperpriors: $\bar{\jmb{y} }_1$ is now modeled by a fully factorized entropy, and $\bar{\jmb{y} }_2$ is modeled by a GMM that only depends on $\bar{\jmb{y} }_1$. We train CE0, both with and without DispSkip, to analyze metrics when no hyperprior side information is used during encoding. Our final stereo model consists of DispSkip, CE0, and the hyperprior, and we showcase that curve as well. 

As shown in Fig.~\ref{fig:abl_results_figure}, DispSkip on its own provides greater gains against the pure IE model at lower bitrates and converges to single-image compression at higher bitrates. In the meantime, CE0 consistently provides performance gains across all bitrates against IE; the improvement is roughly equivalent to that of fitting the Ball\'e hyperprior entropy model for both image codes. When combined with DispSkip (DispSkip + CE0), this model marginally outperforms the Ball\'e hyperprior model. Finally, DispSkip + CE0 + hyperprior (forming our full stereo model) provides the greatest metrics gains across all bitrates.

We observe some cannibalization of additional gains when combining DispSkip, CE0, and the hyperprior. The reduction in gains when we combine DispSkip + CE0 makes intuitive sense: our disparity-warped skip connections focus on reusing redundant information and hence reducing the correlation between the image codes, whereas the entropy of our CE0 model is lower when the correlation in both image codes is higher. Moreover, fitting hyperprior side information that can already help reduce the entropy of a single image may somewhat reduce the additional entropy reduction CE0 can provide. 

\begin{table}[]
	\begin{tabular}{|l|l|l|l|}
		\hline
		Resolution      & 960 x 300       & 480 x 300       & 240 x 150       \\ \hline
		Bitrate (Ours)  & \textbf{0.361}  & \textbf{0.406}  & \textbf{0.437}  \\ 
		MS-SSIM (Ours)  & \textbf{0.9936} & \textbf{0.9935} & \textbf{0.9915} \\ 
		PSNR (Ours)     & \textbf{40.12}  & \textbf{38.22}  & \textbf{35.61}  \\ \hline
		Bitrate (Ball\'e) & 0.414           & 0.479          & 0.518           \\ 
		MS-SSIM (Ball\'e) & 0.9927           & 0.9919          & 0.9905           \\ 
		PSNR (Ball\'e)    & 39.93           & 37.57          & 35.42          \\ \hline
	\end{tabular}
	\vspace{-2mm}
	\caption{Analysis of our stereo compression performance on different camera baseline widths/resolutions from NorthAmerica.}
	\label{tab:baseline_table}
	\vspace{-4mm}
	\end{table}

\vspace{-4pt}
\paragraph{Qualitative Results:} A qualitative demonstration of our model on a stereo pair is given in Fig. \ref{fig:big_qual_results}. We show that our approach contains better overall visual quality at a lower bitrate compared to the Ball\'e model and other codecs. More specifically, our stereo model better preserves overall edges and colors without introducing artifacts. While BPG is competitive with our model on Cityscapes, we observe that BPG magnifies certain high frequency details while distorting lower frequency regions. We leave additional qualitative analysis to supplementary material.
\vspace{-4pt}
\paragraph{Effect of Different Baseline Widths}
To more concretely analyze the impact of the baseline width on our stereo compression model, we recreate copies of NorthAmerica at different resolutions: $960 \times 480$, $480 \times 300$, and $240 \times 150$, with baseline widths of 0.175 m, 0.088 m, 0.044 m respectively. We train with $C=64$ for the highest resolution and $C=32$ for the others. As shown in Table \ref{tab:baseline_table}, we achieve bitrate reductions while increasing perceptual metrics for all resolution levels in NorthAmerica.

\begin{table}[]
	\centering
	\begin{tabular}{|l|l|l|}
		\hline
		& 480 x 300 Res. & 1920 x 720 Res.  \\
		\hline
		Ball\'e & 36 GFlops & 345 GFlops \\
		\hline
		Ours & 141 GFlops & 2700 GFlops \\
		\hline
	\end{tabular}
		\vspace{-2mm}
	\caption{Analysis of FLOPs of our approach compared to Ball\'e.}
	\label{tab:flops_table}
	\vspace{-4mm}
\end{table}

\vspace{-4pt}
\paragraph{Runtime:}

On a GTX 1080-Ti, our stereo model takes 130ms for a 480x300 NorthAmerica pair, and 2246 ms for  a1920x720 Cityscapes pair. Additionally FLOPS are shown in Tab. \ref{tab:flops_table}. 
Range coding is $\mathcal{O}(N)$ in encoding and $\mathcal{O}(N \log n)$ in decoding, where $N$ is \# symbols and $n$ is \# unique symbols.  Our complexity is dominated by the computation of the cost volumes. We note that we can attempt sparse approximations of the cost volume, or different distributional parametrizations instead of a dense softmax to save compute/memory for future work.
\vspace{-4pt}
\paragraph{Disparity Volume:}
To interpret the information learned in our DispSkip connections, Fig. \ref{fig:disp_vis} shows a visualization of our disparity volumes in the encoder/decoder at one particular bitrate level (bitrate ~0.442), for a Cityscapes stereo pair. These visualizations are generated by taking the mode over the probability vector for each disparity dimension in each volume.
The learned disparity maps capture different information at each level, helping to support our justification for predicting a separate disparity volume at each level of the encoder and decoder.

\newcommand{\cityfiga}[1]{
	\begin{overpic}[width=\imw]{#1}
		\put(52,6){ \color{green} \framebox(23, 13){}} 
		\put(0,0){\setlength{\fboxsep}{0pt}\color{orange}\fbox{\includegraphics[viewport=255 30 367 93, clip, height=32pt]{#1}}}
\end{overpic}}
\newcommand{\cityfigb}[1]{
	\begin{overpic}[width=\imw]{#1}
		\put(54,6){ \color{green} \framebox(23, 13){}} 
		\put(0,0){\setlength{\fboxsep}{0pt}\color{orange}\fbox{\includegraphics[viewport=264 30 377 93, clip, height=32pt]{#1}}}

\end{overpic}}
\newcommand{\torfiga}[1]{
	\begin{overpic}[width=\imw]{#1}
		\put(77,15){ \color{green} \framebox(19, 16){}} 
		\put(0,0){\setlength{\fboxsep}{0pt}\color{orange}\fbox{\includegraphics[viewport=377 74 470 151, clip, width=56pt]{#1}}}
\end{overpic}}
\newcommand{\torfigb}[1]{
	\begin{overpic}[width=\imw]{#1}
		\put(79,15){ \color{green} \framebox(19, 16){}} 
		\put(0,0){\setlength{\fboxsep}{0pt}\color{orange}\fbox{\includegraphics[viewport=387 74 480 151, clip, width=56pt]{#1}}}
\end{overpic}}

\section{Conclusion}
We propose a novel deep stereo image compression algorithm, which exploits the content redundancy between the stereo pair to reduce the joint bitrate.
Towards this goal, we propose parameteric skip functions and a conditional entropy model to model the dependence between the images. 
We validate the effectiveness of our method over two large-scale datasets and demostrate that our stereo model reduces the bitrate in the second image by 10-50\% from high to low bitrates, compared to a single-image deep compression model.  
Additionally, we demonstrate that both our skip functions and conditional entropy contribute meaningfully to improving the bitrate and perceptual quality. In the future, we plan to extend our approach to the multi-view image and video compression settings. 


{\small
\bibliographystyle{ieee_fullname}
\bibliography{egbib}
}

\clearpage 
\appendix

\section{Additional Qualitative Results} 

We showcase additional qualitative results comparing our stereo compression model against other baselines, from BPG and JPEG2000 to single-image Ball\'e to our residual coding baseline. We note that for the baselines, we report a separate bitrate per camera, since each camera image is compressed as a single image. However, for our stereo model, we report the joint bitrate divided by 2. The reason for this is that even though our models outputs a separate code for each image, the first code $\bar{\jmb{y}}_1$ may contain additional information to help the compression of the second code $\bar{\jmb{y}}_2$, since it is used as an input for both our skip functions and conditional entropy. We report separate perceptual metrics per camera image for all models. 

\subsection{Additional Qualitative Results from Stereo Model}
\newcommand{\jcaptionc}[1]{\put(0,35.7){\small \colorbox{gray}{\color{white} #1}}}
\newcommand{\cityfigba}[2]{
	\begin{overpic}[width=0.95\linewidth]{#1}
		\put(27,12){ \color{green} \framebox(7, 6){}} 
		\put(34,23){ \color{red} \framebox(3, 3){}} 
		\put(0,0){\setlength{\fboxsep}{0pt}\color{red}\fbox{\includegraphics[viewport=666 444 715 490, clip, height=100pt]{#1}}}
		\put(67.8,0){\setlength{\fboxsep}{0pt}\color{green}\fbox{\includegraphics[viewport=529 235 666 353, clip, height=130pt]{#1}}}
		\jcaptionc{#2}
\end{overpic}}

\newcommand{\jcaptiond}[1]{\put(0,2){\small \colorbox{gray}{\color{white} #1}}}
\newcommand{\tora}[2]{
	\begin{overpic}[width=\imw]{#1}
		\put(74,12){ \color{green} \framebox(23, 20){}} 
		\put(0,0){\setlength{\fboxsep}{0pt}\color{green}\fbox{\includegraphics[viewport=352 59 470 157, clip, height=130pt]{#1}}}
		\jcaptiond{#2}
\end{overpic}}
\newcommand{\torb}[2]{
	\begin{overpic}[width=\imw]{#1}
		\put(77,12){ \color{green} \framebox(23, 20){}} 
		\put(0,0){\setlength{\fboxsep}{0pt}\color{green}\fbox{\includegraphics[viewport=368 59 480 157, clip, height=130pt]{#1}}}
		\jcaptiond{#2}
\end{overpic}}

\begin{figure*} [!htb]
	\centering
	\def\imw{0.5\textwidth}
	\setlength{\tabcolsep}{1pt}
	\begin{tabular}{cc}
		
		\tora{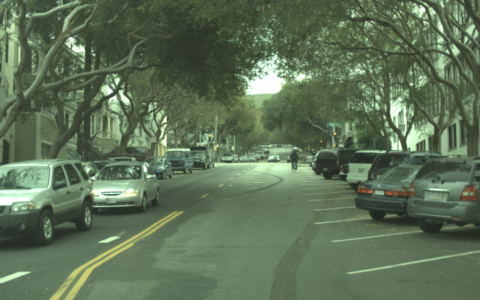}{NorthAmerica (Cam 1)} &
		\torb{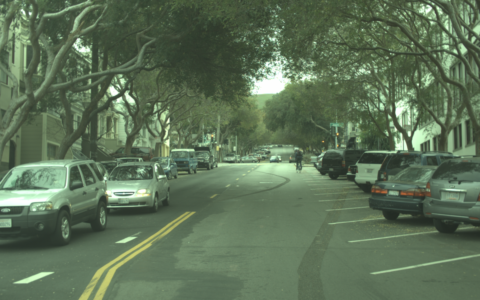}{NorthAmerica (Cam 2)}
		\\ [-1pt]
		
		\tora{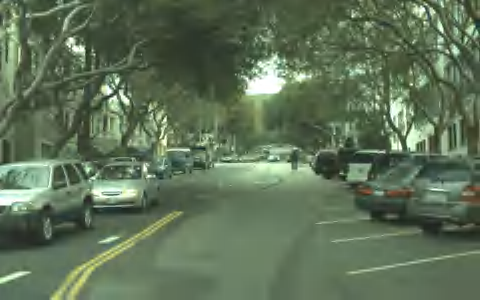}{JPG (Cam 1), Bitrate: 0.343, PSNR: 29.46} &
		\torb{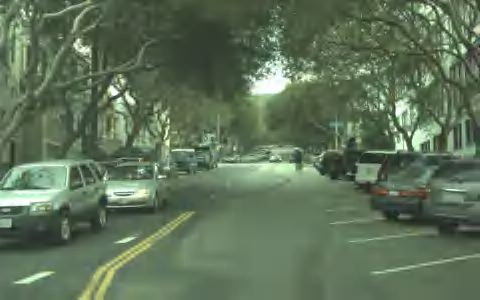}{JPG (Cam 2), Bitrate: 0.339, PSNR: 29.48}
		\\ [-1pt]
		
		\tora{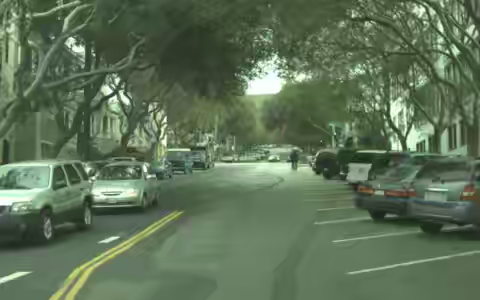}{BPG (Cam 1), Bitrate: 0.331, PSNR: 30.90} &
		\torb{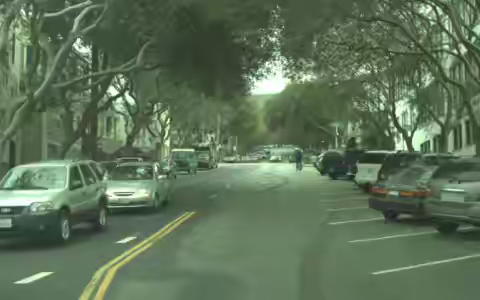}{BPG (Cam 2), Bitrate: 0.332, PSNR: 30.92}
		\\ [-1pt]
		
	\end{tabular}
	\caption{Comparison between the reconstructions of competing baselines and our method on a NorthAmerica stereo pair. We observe that our method yields the highest PSNR at the lowest bitrate compared to all competing methods (34\% reduction in residual bitrate compared to Ball\'e).}
	\label{fig:qual_eval_2}
\end{figure*}

\begin{figure*} [!htb] \ContinuedFloat
	\centering
	\def\imw{0.5\textwidth}
	\setlength{\tabcolsep}{1pt}
	\begin{tabular}{cc}
		
		\tora{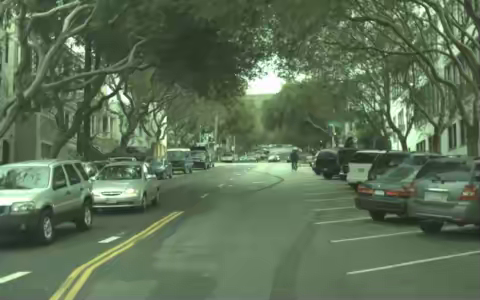}{MV-HEVC (Cam 1), Bitrate: 0.509, PSNR: 32.84} &
		\torb{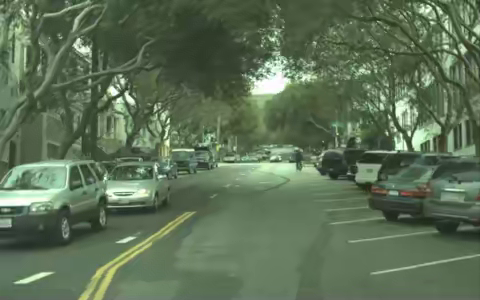}{MV-HEVC (Cam 2), Bitrate: 0.187, PSNR: 31.45}
		\\ [-1pt]
		
		\tora{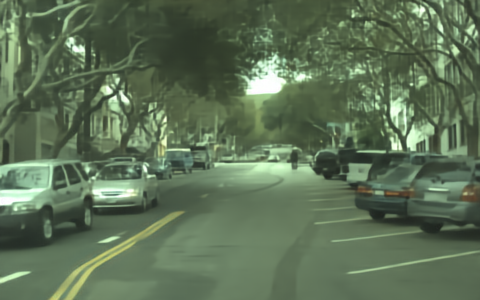}{Ball\'e (Cam 1), Bitrate: 0.376, PSNR: 32.12} &
		\torb{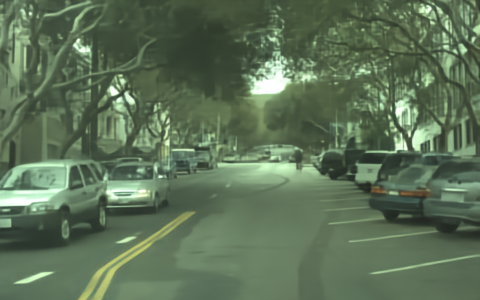}{Ball\'e (Cam 2), Bitrate: 0.372, PSNR: 32.04}
		\\ [-1pt]
		
		\tora{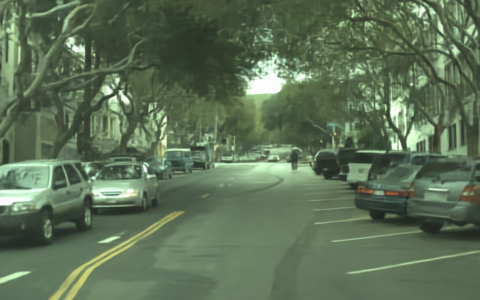}{Ours (Cam 1), Bitrate: 0.369, PSNR: 32.52} &
		\torb{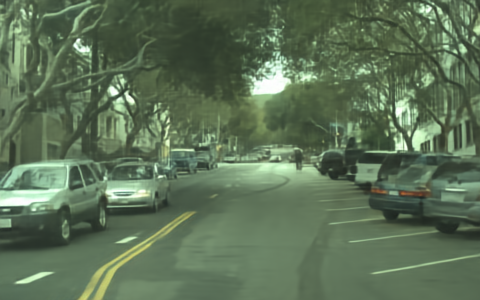}{Ours (Cam 2), Bitrate: 0.251, PSNR: 32.08}
		\\ [-1pt]
		
	\end{tabular}
	\caption{(Continued)}
	\label{fig:qual_eval_2}
\end{figure*}

\newcommand{\torba}[2]{
	\begin{overpic}[width=\imw]{#1}
		\put(0,15){ \color{green} \framebox(20, 20){}} 
		\put(51.5,0){\setlength{\fboxsep}{0pt}\color{green}\fbox{\includegraphics[viewport=0 74 98 172, clip, height=120pt]{#1}}}
		\jcaptiond{#2}
\end{overpic}}
\newcommand{\torbb}[2]{
	\begin{overpic}[width=\imw]{#1}
		\put(0,15){ \color{green} \framebox(20, 20){}} 
		\put(51.5,0){\setlength{\fboxsep}{0pt}\color{green}\fbox{\includegraphics[viewport=0 74 98 172, clip, height=120pt]{#1}}}
		\jcaptiond{#2}
\end{overpic}}

Here, we showcase additional qualitative results for NorthAmerica in \ref{fig:qual_eval_2}, where the image resolution is $480 \times 300$. The closest competing algorithms are generally BPG and the Ball\'e model. Yet on NorthAmerica, our model demonstrates more crisp results at lower bitrates compared to all competing algorithms, as shown in Fig.~\ref{fig:qual_eval_2}. 

\pagebreak
\subsection{Artifacts from Residual Coding Baseline}

We showcase results from our residual coding baseline. In Fig.~\ref{fig:res_coding}, we compare the reconstructions between camera 1 (effectively produced via a single-image Ball\'e network), and camera 2 (produced via the output of motion-compensation using SGM and residual coding using a separate Ball\'e network) on a stereo image pair in Cityscapes. We additionally include the same output from our stereo model. The camera 2 reconstruction has overall higher perceptual metrics in terms of PSNR/MS-SSIM at a lower bitrate, and also that certain regions in the image look undeniably sharper than in camera 1 and in the outputs from our own stereo model (shown by the green boxes). However, we highlight other regions, shown by the red boxes, where there exist jarring artifacts in the camera 2 reconstruction that are not present in camera 1 nor in our stereo model outputs. There are cuts/tears around the boundaries where SGM does not output valid disparities; moreover there exist significant warping artifacts around regions with larger disparities that are predicted less accurately. 

We did not attempt any additional fine-tuning or refinement after merging the residual image with the disparity-warped first reconstruction to construct the second reconstruction. We leave that as an interesting direction to explore in future work. We also note that the artifacts start to go away at higher bitrates, but at that point the overall performance of the stereo residual baseline also deterioriates to below the curve of the single-image Ball\'e model. 

\begin{figure*} [!htb]
	\centering
	\includegraphics[width=\linewidth]{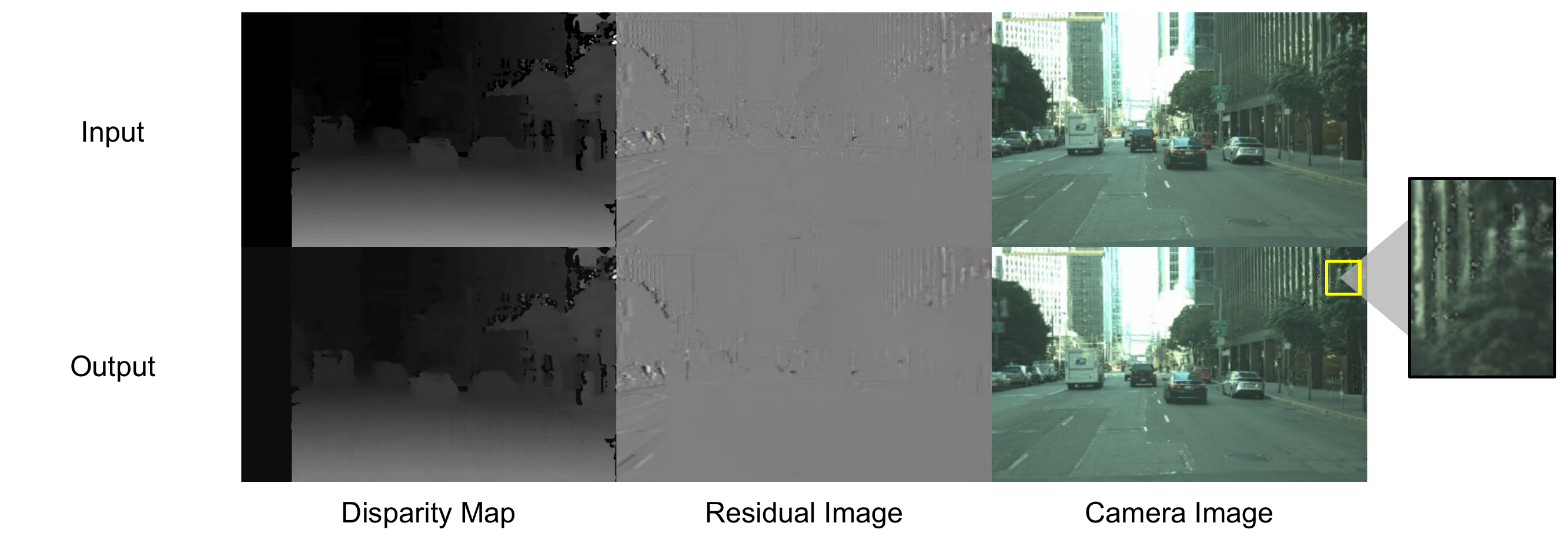}
	\caption{Input/Output disparity map, residual image, and camera image for a sample stereo pair on NorthAmerica (using our stereo residual coding baseline).}
	\label{fig:res_coding_2}
\end{figure*}

\newcommand{\jcaptione}[1]{\put(0,1){\small \colorbox{gray}{\color{white} #1}}}
\newcommand{\cityfigda}[2]{
	\begin{overpic}[width=0.77\linewidth]{#1}
		\put(55,10){ \color{green} \framebox(22, 15){}} 
		\put(11,5){ \color{red} \framebox(13, 30){}} 
		\jcaptione{#2}
\end{overpic}}
\newcommand{\cityfigdb}[2]{
	\begin{overpic}[width=0.77\linewidth]{#1}
		\put(56,10){ \color{green} \framebox(22, 15){}} 
		\put(12,5){ \color{red} \framebox(13, 30){}} 
		\put(0,5){ \color{red} \framebox(8, 20){}} 
		\jcaptione{#2}
\end{overpic}}

\begin{figure*}[!htb]
	\centering
	\cityfigda{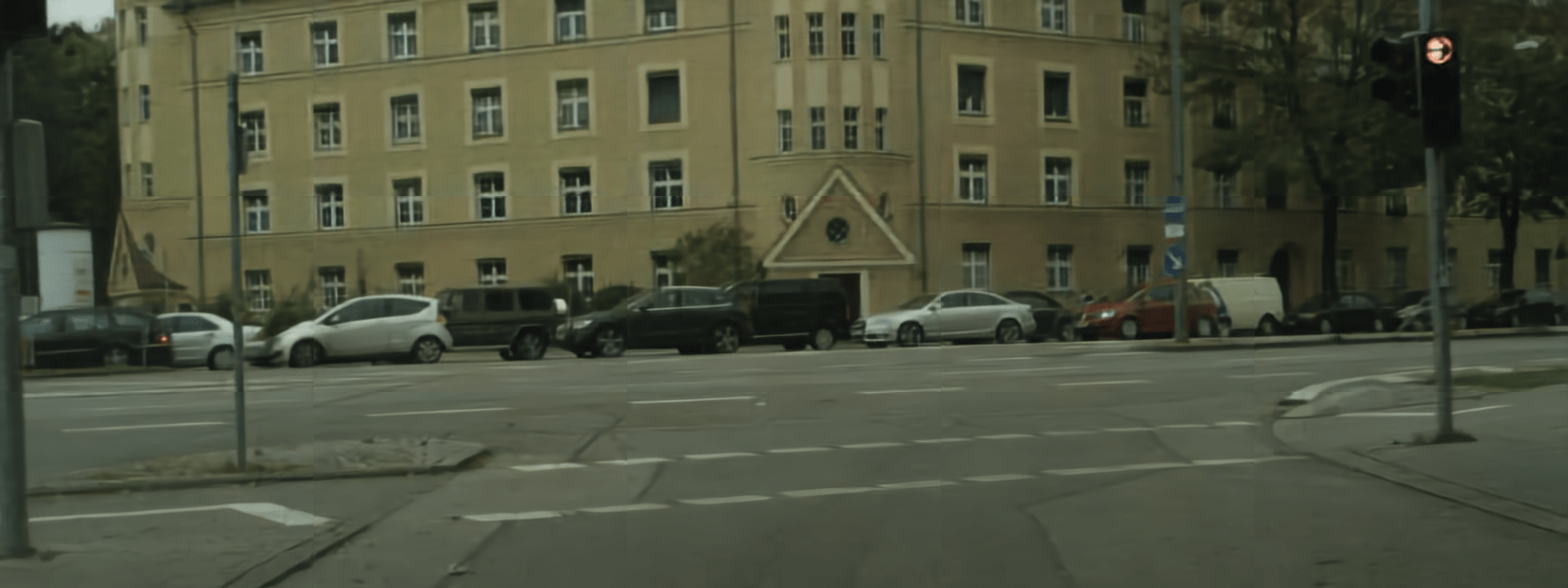}{Deep Residual Coding (Camera 1), Bitrate: 0.077, PSNR: 34.89}
	\cityfigdb{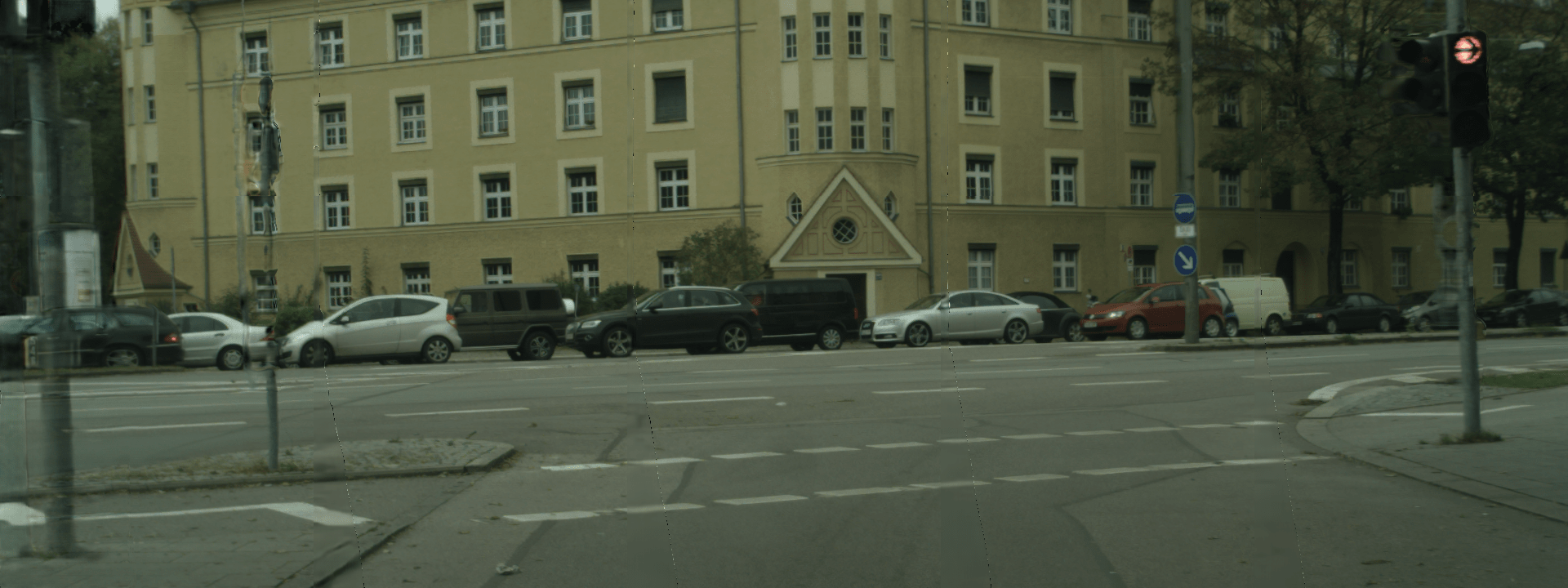}{Deep Residual Coding (Camera 2), Bitrate: 0.032, PSNR: 35.95}
	\cityfigda{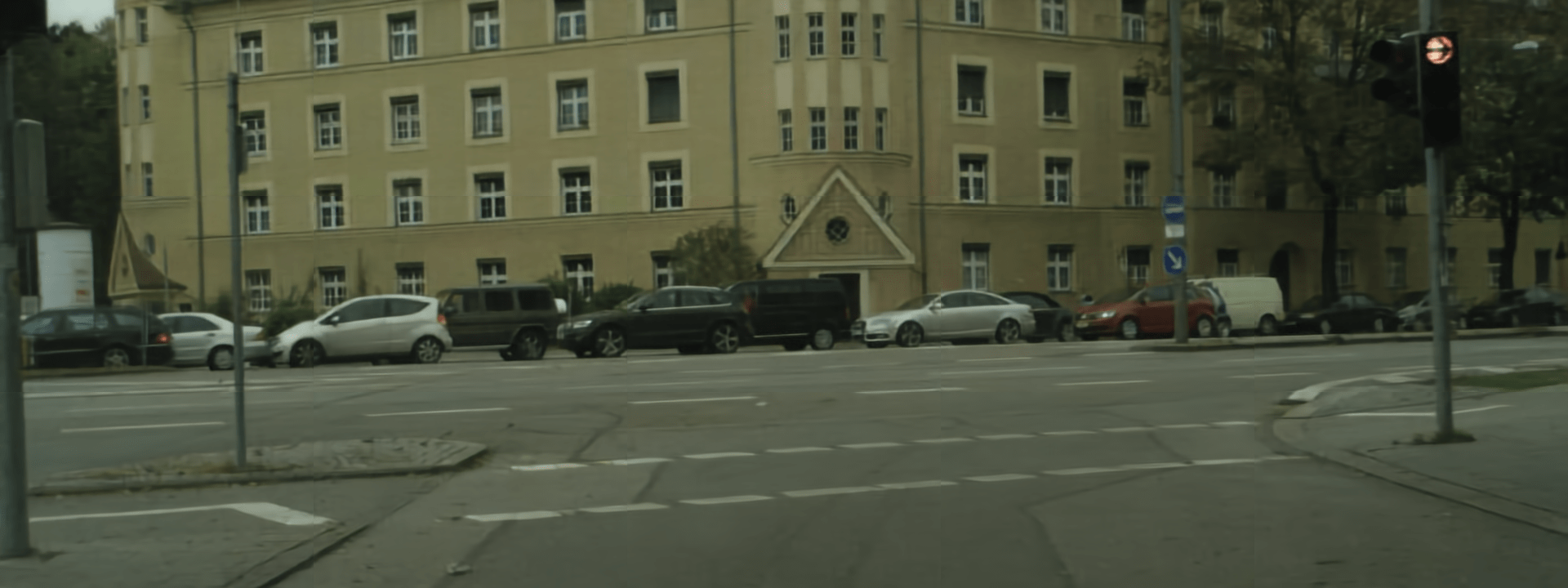}{Ours (Camera 1), Bitrate: 0.099, PSNR: 36.73}
	\cityfigdb{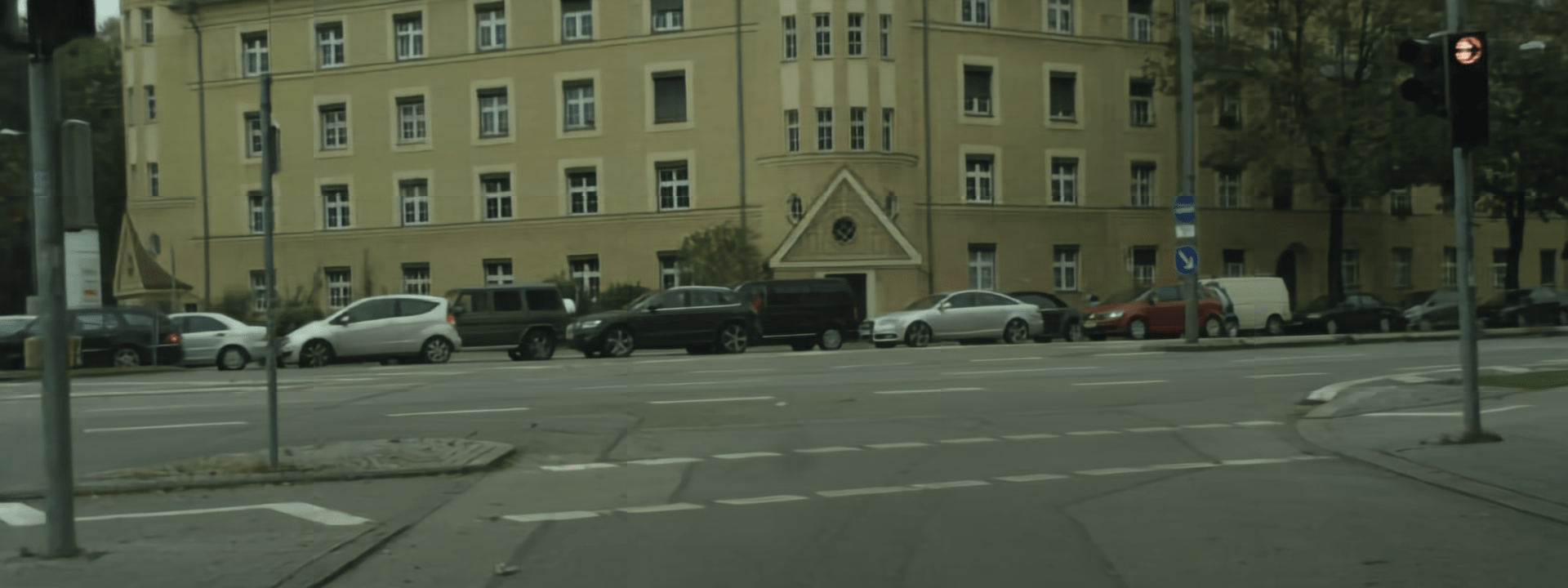}{Ours (Camera 2), Bitrate: 0.024, PSNR: 35.80}
	\caption{Comparison between the reconstructions from the two cameras using our deep residual coding baseline as well as our stereo model, for a Cityscapes stereo pair. The green box demonstrates where our residual baseline reconstruction (Cam 2) has sharper image quality than our stereo model. The red boxes demonstrate where the residual baseline reconstruction (Cam 2) introduces artifacts that are absent in our stereo model. }
	\label{fig:res_coding}
\end{figure*}

The artifacts also exist in NorthAmerica, where our deep residual coding baseline underperforms even single-image compression at all bitrates. We show a sample original/reconstructed disparity map, residual image, and final image in Fig. \ref{fig:res_coding_2}.

\pagebreak
\section{Expanded Ablation Studies}

In our ablation study in the main paper (see Fig 5. in the main paper), we measure the independent effects of our parametric skip functions, conditional entropy, and hyperprior by adding them on top of a factorized-prior model. Here, in our expanded ablation study, we measure the separate effects of our parametric skip functions and conditional entropy when combined with a single-image hyperprior model. 

The results are shown in Fig. \ref{fig:abl_results_figure}. We observe that our results follow a similar trend to that in the main paper. DispSkip provides the highest bitrate savings and perceptual metrics gains at lower bitrates, and decreases for higher bitrates (especially for Cityscapes). Meanwhile, adding a conditional entropy component adds relatively consistent bitrate savings at all levels compared to the hyperprior model. We also observe cannibalization effects when combining DispSkip with conditional entropy, which is also observed in the main paper; yet again, combining DispSkip with conditional entropy yields the best results at all bitrates. 
\begin{figure*} [!htb]
	\centering
	\includegraphics[height=0.3\linewidth]{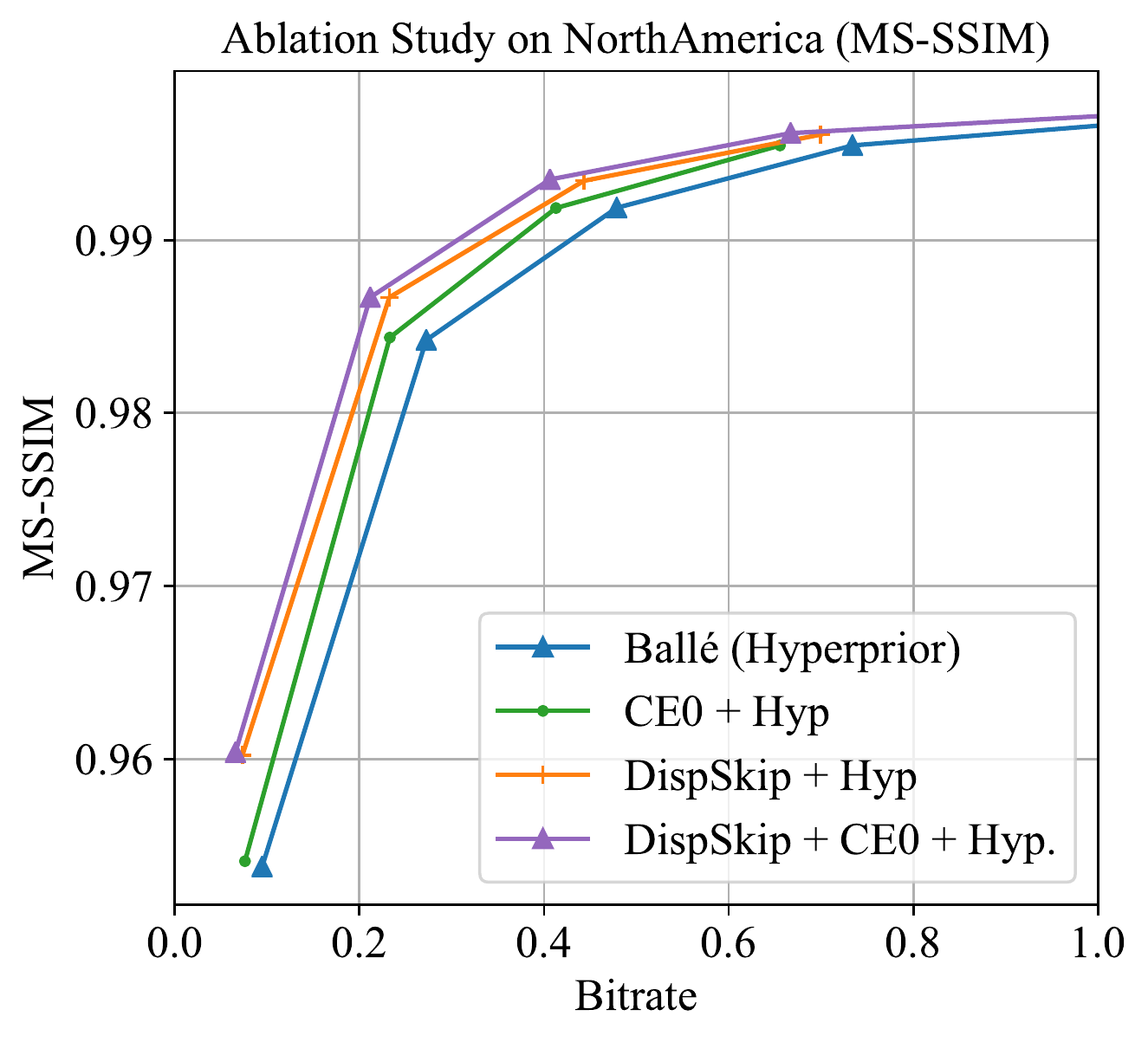}
	\includegraphics[height=0.3\linewidth]{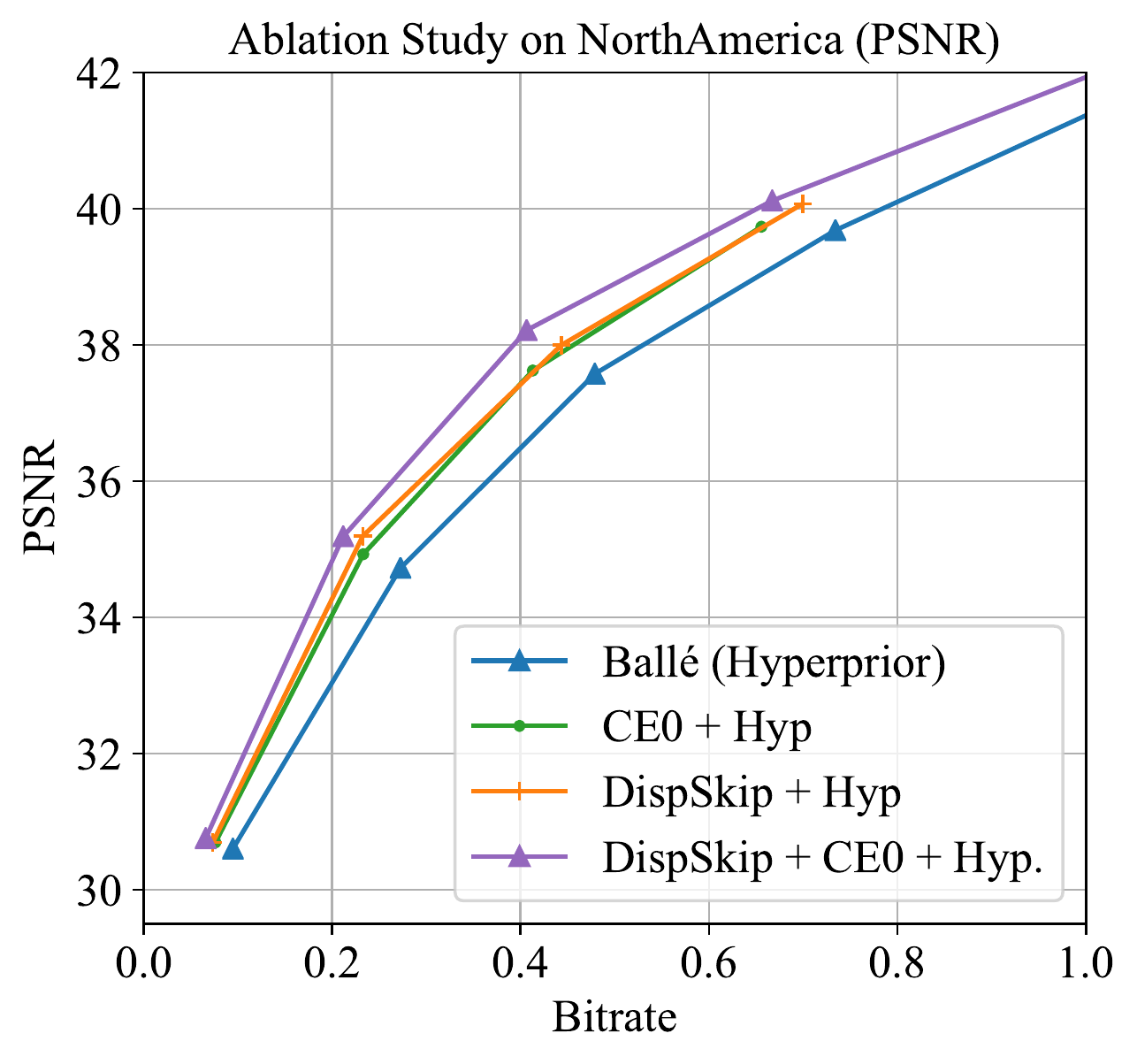}
	\includegraphics[height=0.3\linewidth]{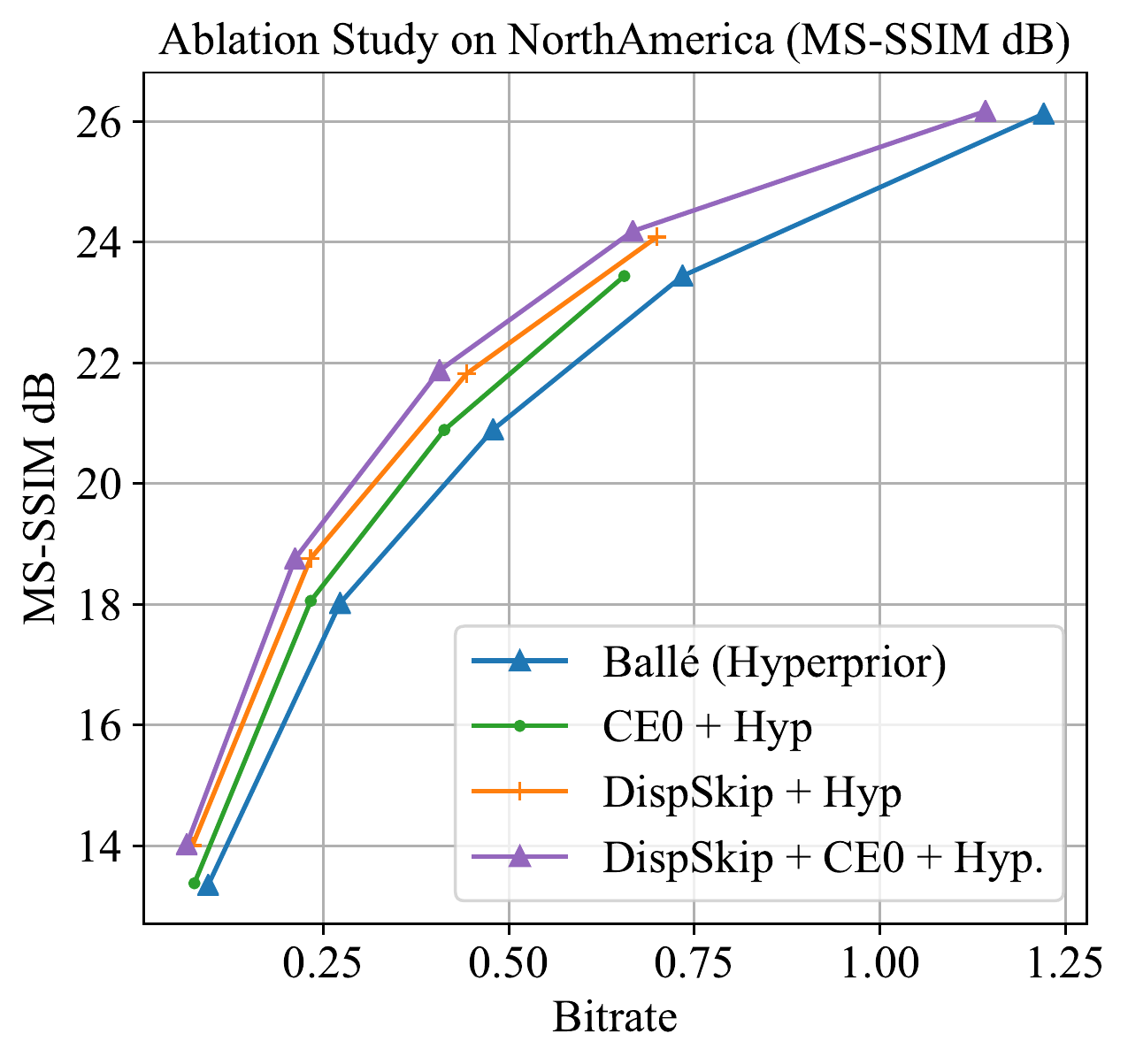} \\
	\includegraphics[height=0.3\linewidth]{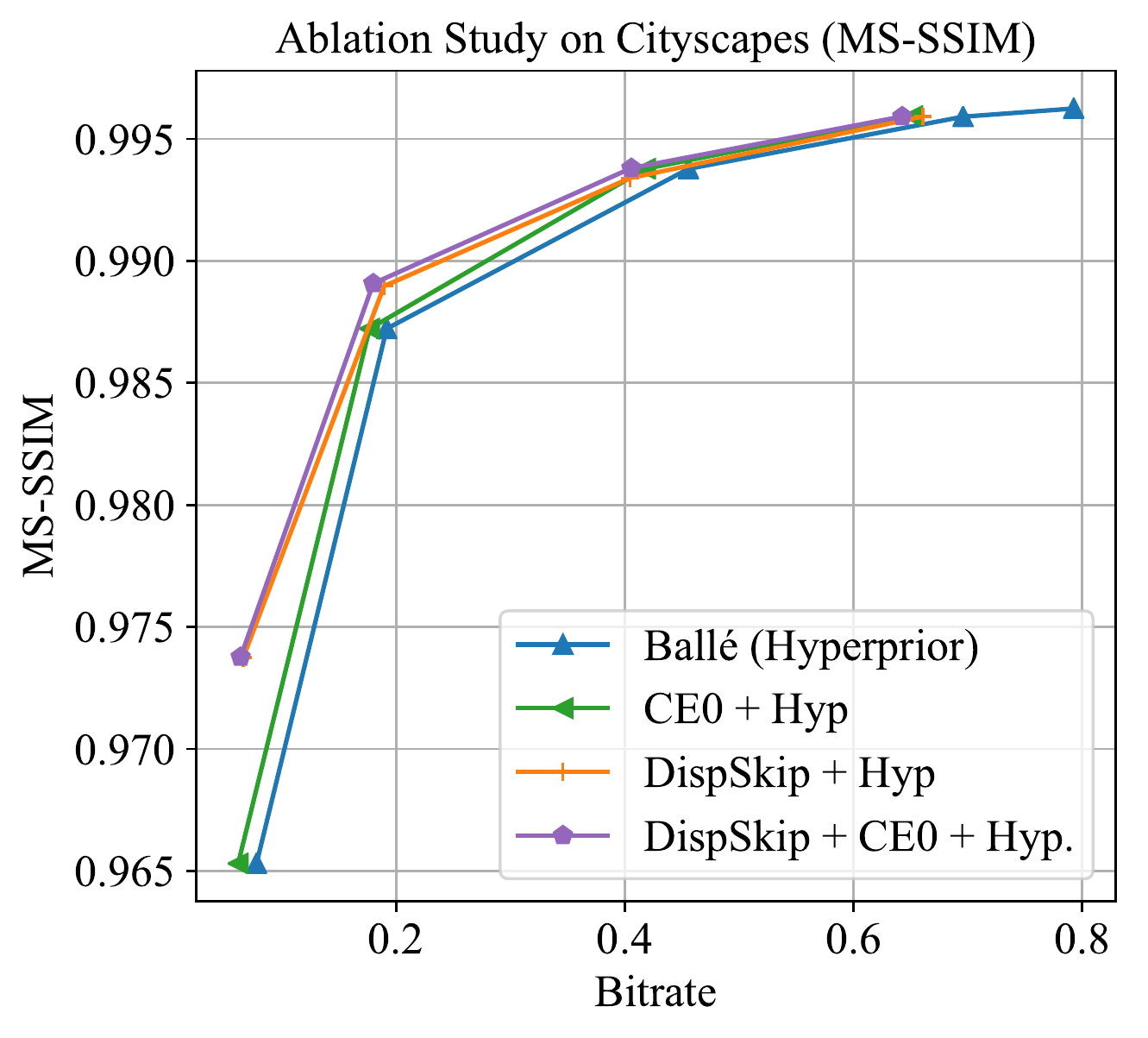}
	\includegraphics[height=0.3\linewidth]{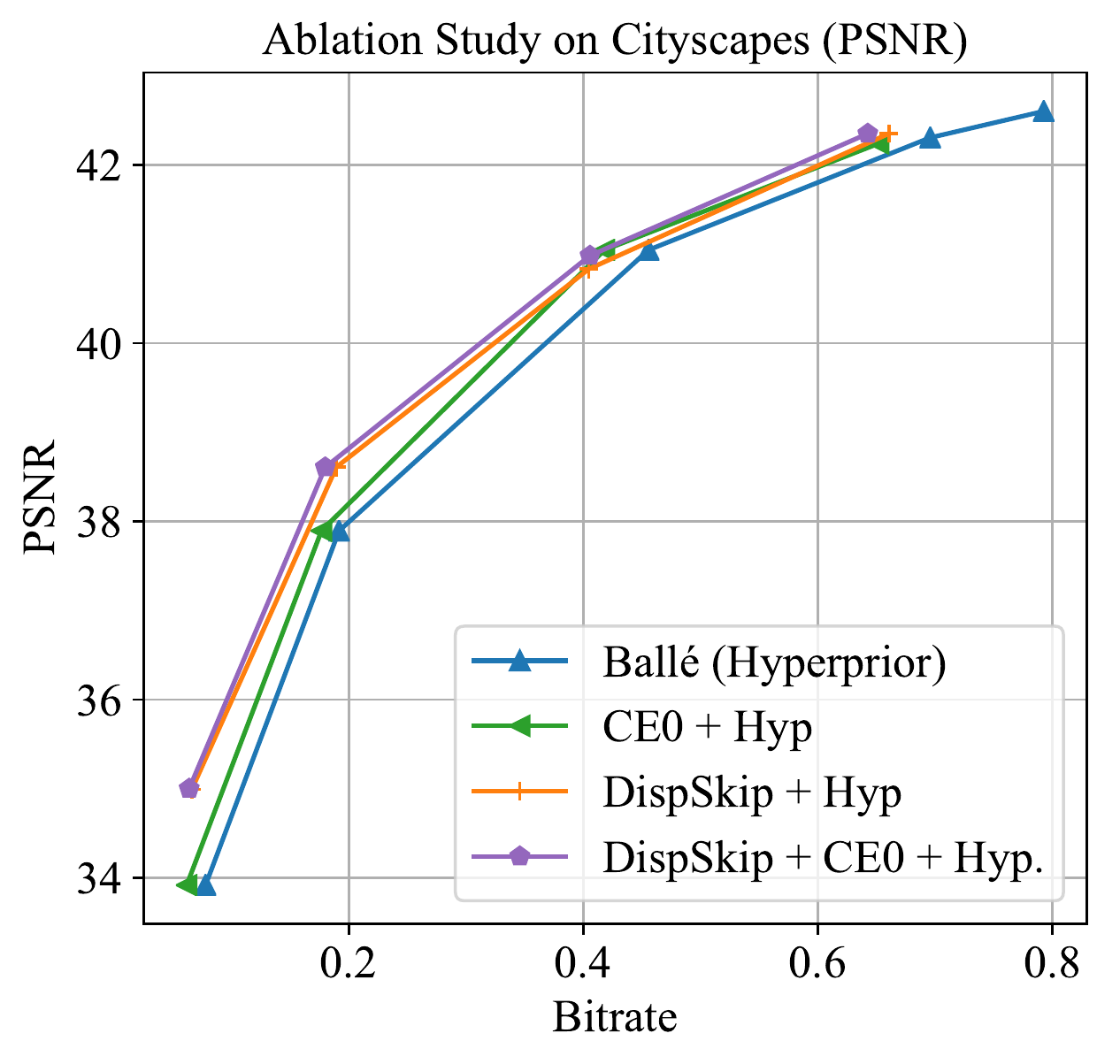}
	\includegraphics[height=0.3\linewidth]{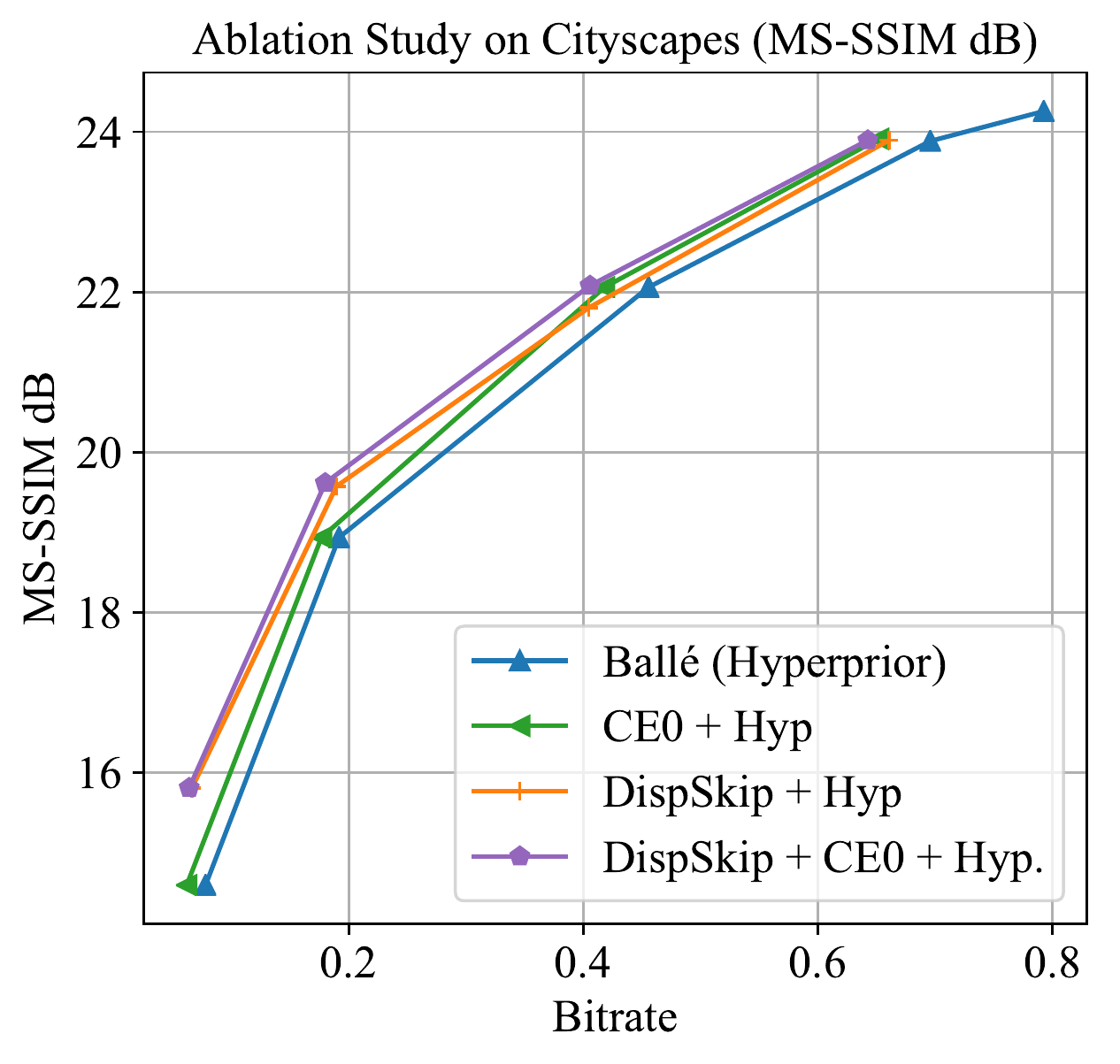}
	\caption{Additional ablation study. For both datasets, we analyze the independent and combined effects of our skip functions (DispSkip) and the conditional entropy on top of the single-image hyperprior model.}
	\label{fig:abl_results_figure}
\end{figure*}

\pagebreak
\section{Additional Architecture Details} \label{sec:arch_supp}

We provide additional architecture details in this section on various aspects of our model. First, in Section \ref{sec:arch_1}, we provide some more details about our main encoder/decoder architecture. Then, in Section \ref{sec:arch_2}, we provide architecture details about the main components of our parametric skip functions: predicting the global context, predicting the cost volume at each level of the encoder/decoder, as well as the final feature aggregation. Finally, in Section \ref{sec:arch_3}, we provide details for the varous components that make up our conditional entropy model: our hyper-encoder (deriving hyperpriors from our image code), our factorized prior entropy model for our hyperpriors, and our GMM-based model for our image codes. 

\subsection{Additional Architecture Details for Encoder/Decoder} \label{sec:arch_1}
The number of channels for each intermediate layer in both the encoder/decoder of each image is set to $N$, and the number of channels of each of the two codes, $\bar{\jmb{y}}_1$, $\bar{\jmb{y}}_2$ is set to $M$. 
For the lower bitrates ($< 0.7$), we set $N=100$ and $M=140$; we found that setting a smaller bottleneck didn't affect model performance too much and allowed the models to train much faster. For the higher bitrates ($\geq 0.7$), we set $N=192$ and $M=256$. 

\subsection{Architecture Details of Parametric Skip Function} \label{sec:arch_2}

Recall that our parametric skip functions consist of four main components. A \textbf{global context} feature is predicted from the code of image 1 $\bar{\jmb{y}}_1$, in order to capture global information from image 1. Then, at each level of the encoder/decoder, we predict a \textbf{stereo cost volume} from $\jmb{h}^{t-1}_1$, $\jmb{h}^{t-1}_2$ - the feature maps of image 1 and 2 from the previous layer - as well as the global context feature. We use the cost volume to \textbf{densely warp} $\jmb{h}^{t-1}_1$ from image 1 to image 2, and finally \textbf{aggregate} this warped feature with $\jmb{h}^{t-1}_2$. We describe the architecture details of predicting the global context, predicting the stereo cost volume at each level, and aggregating the features below. 

\paragraph{Global Context:}The global context module takes as input $\bar{\jmb{y}}_1$, the first image code, with dimensions $M \times H/16 \times W/16$, where $M$ is the channel dimension and $H,W$ are the height/width of the original image. It passes $\bar{\jmb{y}}_1$ through four 2D convolutional layers. Each conv layer except the last is followed by a GroupNorm \cite{wu_groupnorm} and ReLU layer. In general we use GroupNorm instead of BatchNorm \cite{ioffe_batchnorm} in our models due to our small batch sizes.  

The dimension of each intermediate feature is $F \cdot C$, where $C$ is our maximum disparity and $F$ is a multiplicative factor. The final global context output after the convolutional layer is $(F \cdot C) \times H/16 \times W/16$, which we reshape into a 4D volume: $F \times C \times H/16 \times W/16$. Hence our global context can be seen as an initial cost volume (with an additional feature dimension), which we will provide as input to our skip functions at each level of our encoder/decoder. 

Note that we have three levels of skip functions in both the encoder/decoder, predicting cost volumes of dimensions $C \times H/2 \times W/2$, $C \times H/4 \times W/4$, and $C \times H/8 \times W/8$ for the encoder and of dimensions $C \times H/8 \times W/8$, $C \times H/4 \times W/4$, and $C \times H/2 \times W/2$ for the decoder. Since the disparity dimension remains fixed regardless of spatial resolution, the lower resolution cost volumes effectively have a greater receptive field than the higher resolution volumes (ideally we would like the higher resolution volumes to have a big receptive field but this is subject to GPU memory limits). This also implies that the disparity dimensions are not spatially aligned across different spatial resolutions nor with our global context (at the lowest spatial resolution $H/16 \times W/16$), so feeding our global context as is to each level doesn't make sense. 

Instead, we ensure that $F$ is divisible by 3, and our global context volume actually represents a concatenation of three "sub" context volumes of dimensions $F_0 \times C \times H/16 \times W/16$, where $F_0 = F/3$. Each sub-context volume is mapped as an input to a skip function at a corresponding resolution level in both the encoder/decoder (so one sub-context volume is mapped to the skip function in both the encoder and decoder at resolution $H/8,W/8$, etc.). This allows each sub-context volume to represent a lower-resolution feature representation to help predict a specific cost volume at a particular resolution level, as opposed to helping predict all cost volumes across all resolution levels. 

\begin{figure*}
	\centering
	\includegraphics[width=\linewidth]{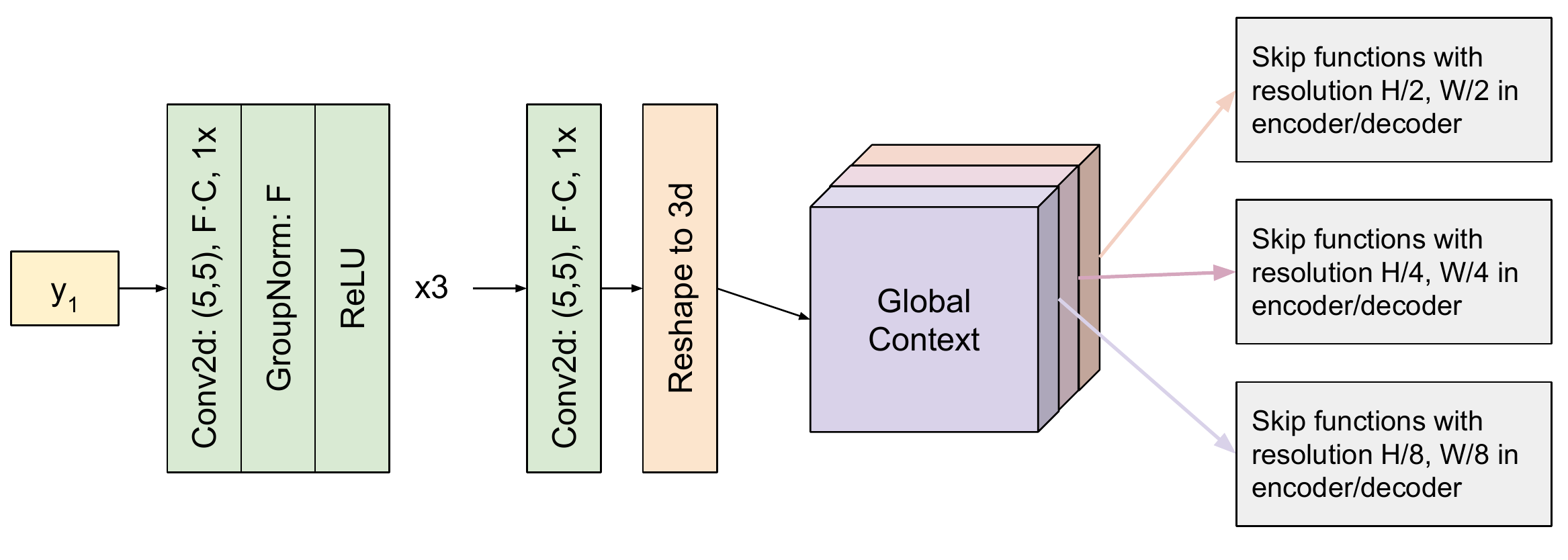}
	\caption{Architecture diagram for producing the global context volume from $\bar{\jmb{y}}_1$, with 3 subcontexts. Each subcontext is passed to the two corresponding skip functions at that resolution level, one in the encoder and one in the decoder.}
	\label{fig:global_context}
\end{figure*}

A network diagram is shown in Fig. \ref{fig:global_context}. We set $F=21$ in our experiments. As mentioned in our experiments, we set $C=32$ for NorthAmerica and $C=64$ for Cityscapes. We set GroupNorm to have $F$ groups, with $C$ channels per group. 

\paragraph{Stereo Cost Volume} 
If the input features to each skip function are at level $t - 1$ with resolution $r$, denote the corresponding sub-context volume from the global context as $\jmb{d}^{r}$. The task of predicting the cost volume used for warping takes in $\jmb{d}^{r}$, as well as $\jmb{h}^{t-1}_1$, $\jmb{h}^{t-1}_2$ as input. 

We concatenate $\jmb{h}^{t-1}_1$, $\jmb{h}^{t-1}_2$ into a $2N \times H^{t-1} \times W^{t-1}$ feature, and feed it through 2 2d convolutions, followed by GroupNorm (with 4 groups per module) and ReLU after each conv. The output feature has dimensions $N \times H^{t-1} \times W^{t-1}$. 

In another branch, we feed $\jmb{d}^{r}$, the sub-context volume, through an upsampling 3d conv. to match the spatial resolution of $\jmb{h}^{t-1}_1$, $\jmb{h}^{t-1}_2$ (which is $H^{t-1}, W^{t-1}$), followed by another 3d conv. Each 3d conv is also followed by GroupNorm (1 group per module) and ReLU, and the intermediate feature channel dimensions are $C \cdot F_0$. The output feature has dimensions $F_0 \times C \times H^{t-1} \times W^{t-1}$, and we collapse this back into a 2d feature representation: $(F_0 \cdot C) \times H^{t-1} \times W^{t-1}$. 

We concatenate the outputs of both feature branches and add 3 more 2d conv layers, with intermediate feature dimension $N$, each except the last followed by GroupNorm(4 groups each) and ReLU. The final cost volume has dimensions $C \times H^{t-1} \times W^{t-1}$, with a softmax layer applied over the disparity dimension for every $0 \leq i,j \leq  H^{t-1},W^{t-1}$. 

\begin{figure*}
	\centering
	\includegraphics[width=\linewidth]{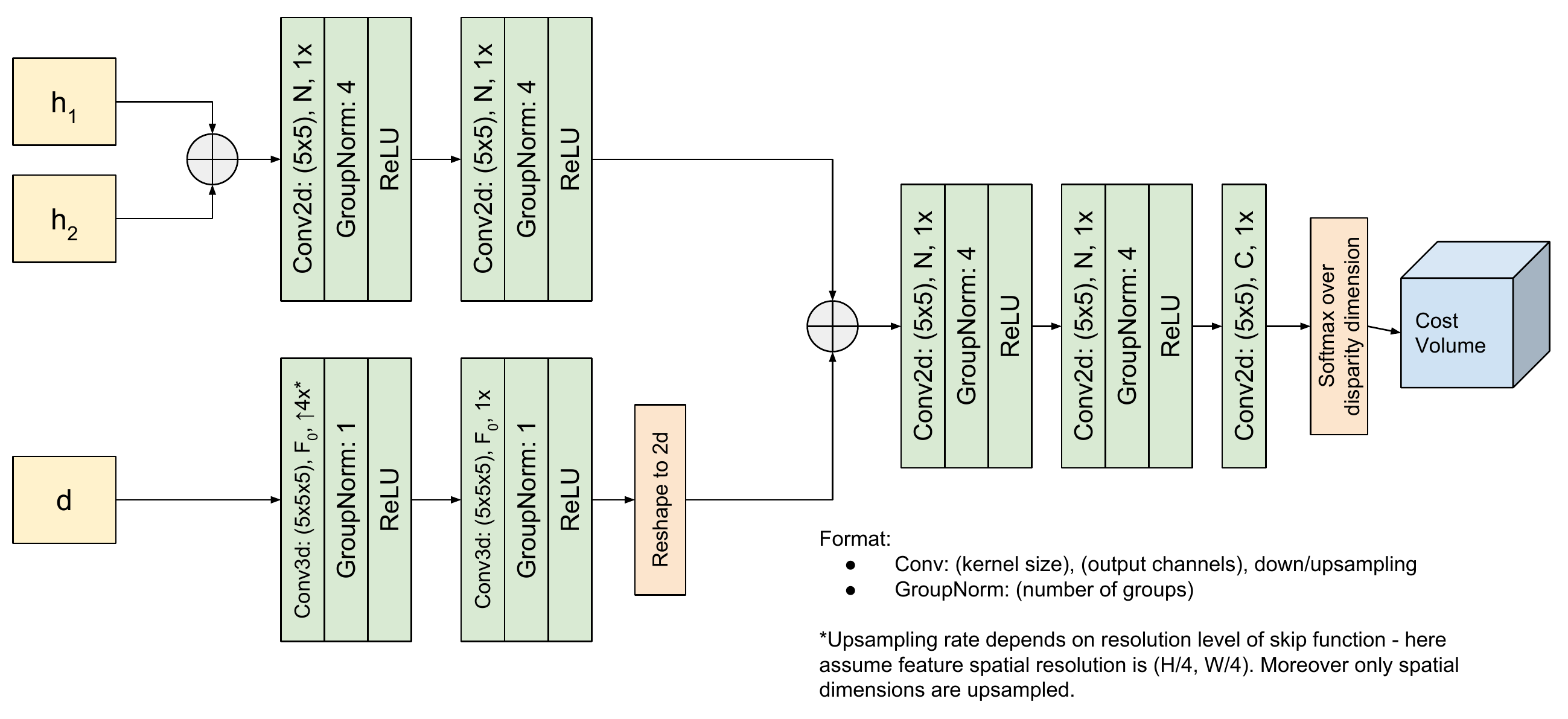}
	\caption{Architecture diagram for producing the cost volume from $\jmb{h}^{t-1}_1$, $\jmb{h}^{t-1}_2$, $\jmb{d}^{r}$. The $\oplus$ symbol represents concatenating two tensors along the channel dimension. }
	\label{fig:cost_volume}
\end{figure*}

A network diagram for predicting the cost volume is given in Fig. \ref{fig:cost_volume}. 

\paragraph{Aggregation Function}
Our aggregation function $\jmb{h}^{t}_2 = a(\jmb{g}^{t-1}_{2}, \jmb{h}^{t-1}_2)$ is fairly simple - since $\jmb{g}^{t-1}_{2}$ and $\jmb{h}^{t-1}_2$ have the same spatial resolution, we concatenate them along the channel dimension. Then we apply a downsampling/upsampling conv as part of the second image's encoder/decoder, as shown in Fig. 1 of the main paper.

\subsection{Architecture Details for Entropy Models} \label{sec:arch_3}
\begin{figure*}[!htb]
	\centering
	\includegraphics[width=\linewidth]{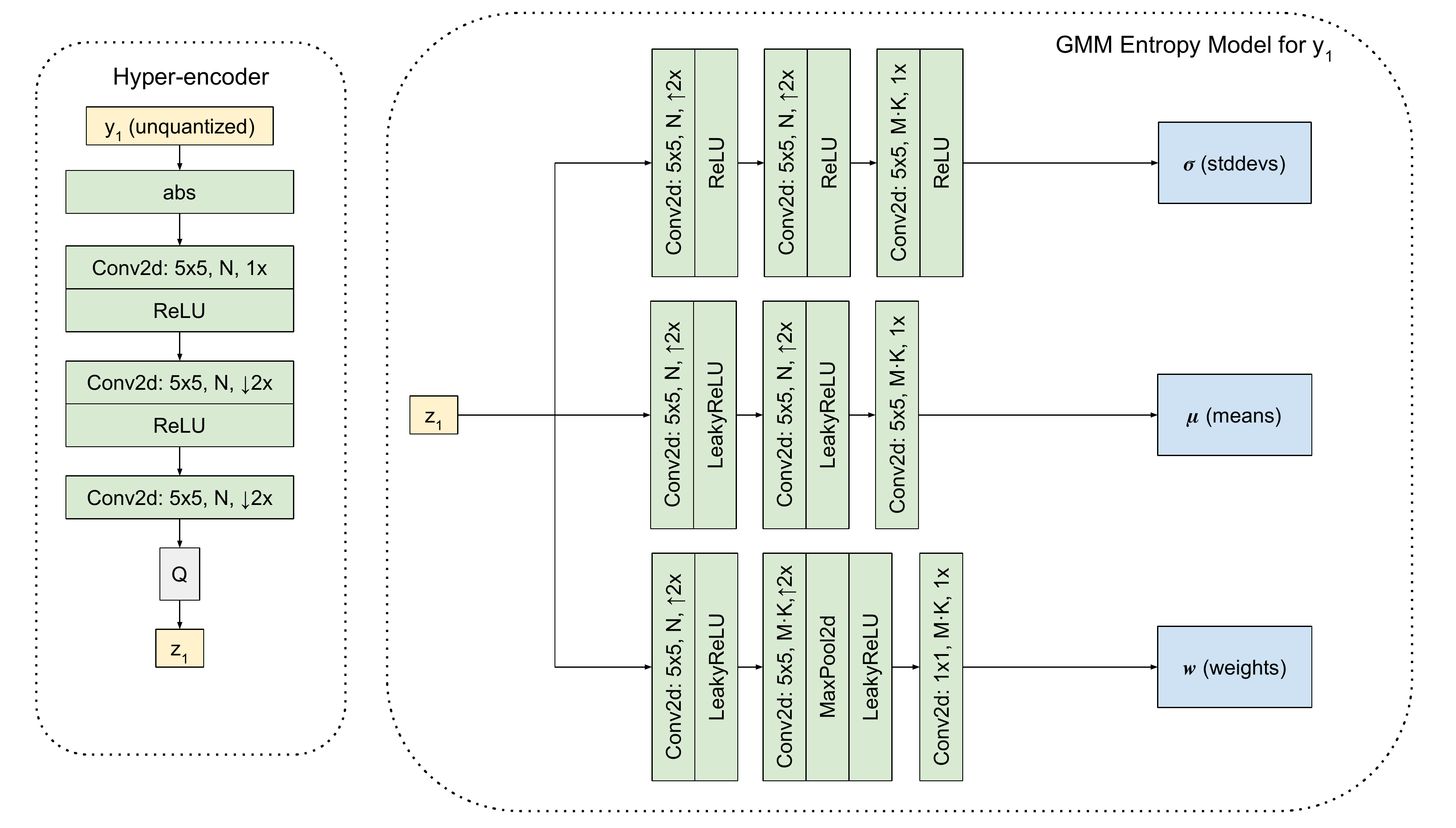}
	\caption{Architecture diagram detailing the hyper-encoder (left) as well as the full entropy model of $\bar{\jmb{y}}_1$ (right). We note that the input to the hyper-encoder is $\jmb{y}_1$ (the continuous representation before being fed to the quantizer), not $\bar{\jmb{y}}_1$ (the noisy representation of  $\jmb{y}_1$ we apply as part of the quantizer during training). The hyperencoder produces $\bar{\jmb{z}}_1$, which we then feed into the GMM entropy model. }
	\label{fig:entropy_model_1}
\end{figure*}

\paragraph{Hyper-encoder:} Our "hyper-encoder" derives the hyperprior variables, $\bar{\jmb{z}}_1,\bar{\jmb{z}}_2$ from $\jmb{y}_1,\jmb{y}_2$. Note that we pass the unquantized continuous representation $\jmb{y}$ into the hyper-encoder, not $\bar{\jmb{y}}$, the noisy representation produced by the quantizer during training. Each $\jmb{y}$ is fed through 3 convolution layers, with ReLUs following the first two and the last two being downampling; then a quantizer is applied to produce $\bar{\jmb{z}}$.  An illustration can be shown in Fig. \ref{fig:entropy_model_1}. 

\paragraph{Hyperprior Entropy Model:}
We follow \cite{balle_varhyperprior} in designing the factorized entropy model for the hyperprior - specifically in modeling $c_{i}(\bar{z}_{i}; \jcb{\theta}_{\bar{\jmb{z}}})$. In order to define a valid cumulative density, $c_{i}(\bar{z}_{i}; \jcb{\theta}_{\bar{\jmb{z}}})$ must map values between $[0,1]$ and be monotonically increasing. The input $\bar{z}_{i}$ and the output must also be univariate (dimension = 1). 

We set $c_i$ to be a two-step nonlinear function as follows: 

\begin{equation}
c_{i}(\bar{z}_{i}; \jcb{\theta}_{\bar{\jmb{z}}}) = f_2 \circ f_1 
\end{equation}

where $f_1: \mathbb{R}^{1} \rightarrow \mathbb{R}^{3}$ and $f_2: \mathbb{R}^{3} \rightarrow \mathbb{R}^{1}$. The nature of each $f_k$ is defined as follows: 
\begin{equation}
\begin{aligned}
f_k(\jmb{x}) &= g_k(\text{softplus}(\jmb{H}^k) \jmb{x} + \jmb{b}^k) \\
g_1(\jmb{x}) &= \jmb{x} + \text{tanh}(\jmb{a}^k) \odot \text{tanh}(\jmb{x}) \\
g_2(\jmb{x}) &= \text{sigmoid}(\jmb{x})
\end{aligned}
\end{equation}

where $\jmb{H}^k$ are matrices, $\jmb{b}^k$ and $\jmb{a}^k$ are vectors, and $\odot$ is elementwise multiplication. This formulation satisfies the conditions to be a valid CDF. For more details and justifications about this manner of designing a factorized prior, see Appendix 6.1 in \cite{balle_varhyperprior}.

We use this same factorized prior formulation for modeling our main image codes in our models without hyperpriors in our ablation study (Section 4.3 in the main paper). For our IE models, we used the factorized prior model for both image codes. For our CE0 models, we used this factorized prior model for the first image code. 

\paragraph{Image Codes Entropy Model:} 
We now describe the GMM-based conditional entropy model for the image codes: $\bar{\jmb{y}}_1,\bar{\jmb{y}}_2$. We start with $\bar{\jmb{y}}_1$. Recall that we define $p_{1, i}(\bar{y}_{1,i} | \bar{\jmb{z}}_1; \jcb{\theta}_{\bar{\jmb{y}}_1}) = (q_{1,i} * u)(\bar{y}_{1,i})$, where $q_{1,i} = \sum_k{w_{ik} \mathcal{N}(\mu_{ik},\sigma^2_{ik})})$.  We predict $\jmb{w}$, $\jmb{\mu}$, and $\jmb{\sigma}$ as functions of $\jmb{z}_1$ given $\jcb{\theta}_{\bar{\jmb{y}}_1}$:  $w(\bar{\jmb{z}}_1; \jcb{\theta}_{\bar{\jmb{y}}_1})$, $\mu(\bar{\jmb{z}}_1; \jcb{\theta}_{\bar{\jmb{y}}_1})$, $\sigma(\bar{\jmb{z}}_1; \jcb{\theta}_{\bar{\jmb{y}}_1})$ - where $\jmb{w}$, $\jmb{\mu}$, and $\jmb{\sigma}$ represent the vectors of all the individual values $w_{ik},\mu_{ik},\sigma_{ik}$. $\jmb{\sigma}$ and $\jmb{\mu}$ have the same spatial resolution as $\bar{\jmb{y}}_1$ with up to $K$ times the number of channels, where $K$ is the number of mixtures ($(M \cdot K) \times H/16 \times W/16$).  Moreover, to reduce the number of parameters and help maintain spatial invariance, we assume that weights are fixed per channel, so weights have dimensions $(M \cdot K) \times 1 \times 1$. The network diagram is shown in Fig. \ref{fig:entropy_model_1}, and there are a few key details per branch. Namely, we apply a ReLU to the last layer of $\sigma(\bar{\jmb{z}}_1; \jcb{\theta}_{\bar{\jmb{y}}_1})$ to keep standard deviations positive. For weights, we apply a pooling layer after the second conv to collapse the spatial dimension, then a softmax per mixture to keep weights normalized. 

\begin{figure*} [!htb]
	\centering
	\includegraphics[width=\linewidth]{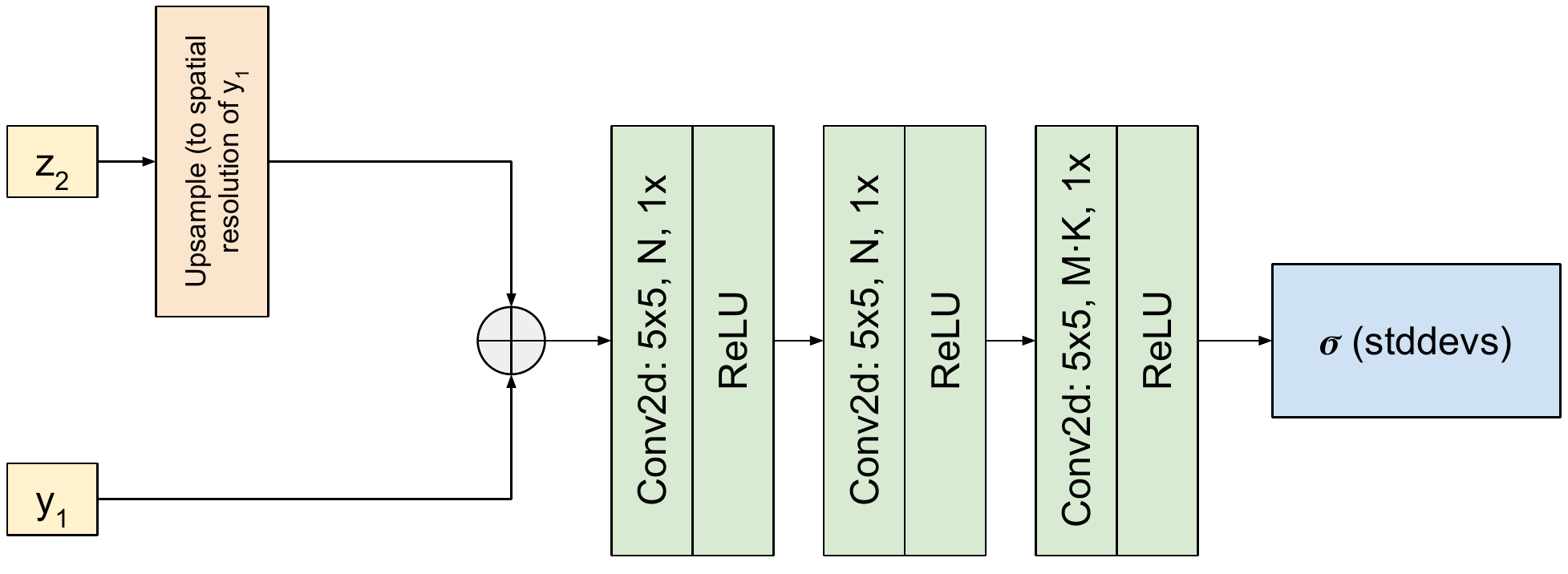}
	\caption{Architecture diagram illustrating how $\jmb{\sigma}$ is predicted for $\bar{\jmb{y}}_2$.  The key difference is that $\bar{\jmb{y}}_1$ is concatenated with an upsampled $\bar{\jmb{z}}_2$ and the convolutions are no longer upsampling. The changes to predict $\jmb{\mu}$, $\jmb{w}$ are the same. }
	\label{fig:entropy_model_2}
\end{figure*}

We follow a similar process to model $p_{2, i}(\bar{y}_{1,i} | \bar{\jmb{z}}_2, \bar{\jmb{y}}_1; \jcb{\theta}_{\bar{\jmb{y}}_2}) = (q_{2,i} * u)(\bar{y}_{2,i})$. However, the network structure for predicting $\jmb{w}$, $\jmb{\mu}$, and $\jmb{\sigma}$ is slightly different because $\bar{\jmb{z}}_2, \bar{\jmb{y}}_1$ are not the same dimension. Instead, we first upsample $\bar{\jmb{z}}_2$ to an intermediate value with the same dimensions of $\bar{\jmb{y}}_1$. Then we can concatenate this intermediate value with $\bar{\jmb{y}}_1$ across the channel dimension and pass it through the convolutions. The convolutions themselves are no longer upsampling, since the input is at the same desired spatial resolution as the output. An example for predicting $\jmb{\sigma}$ is shown in Fig. \ref{fig:entropy_model_2}.

\section{Effect of Different Lossless Coders:} \label{sec:encoding_details}
For lossless encoding, we compare our range coding \cite{martin_range} implementation against Huffman coding and zlib \cite{zlib}.
We find that range coding achieves a bitrate that is within 1-2\% of the Shannon entropy lower bound. As a comparison, our Huffman coding implementation with a tuned chunk size uses 35-50\% more bits than the Shannon entropy. Finally, the DEFLATE algorithm used in zlib (a combination of LZ77 and Huffman) uses between 150\%-200\% more bits.

\end{document}